\documentclass[aps,prd,twocolumn,superscriptaddress,preprintnumbers,nofootinbib,10pt]{revtex4-1}
\pdfoutput=1
\usepackage{color,graphicx,epsfig}
\usepackage{ifpdf}
\usepackage{amsmath}
\usepackage{bm}
\usepackage{color}
\usepackage[english]{babel}
\usepackage{graphicx}
\usepackage{amsfonts}
\usepackage{amssymb}
\usepackage{hyperref}
\usepackage{enumerate}

\definecolor{nicered}{rgb}{0.7,0.1,0.1}
\definecolor{nicegreen}{rgb}{0.1,0.5,0.1}
\hypersetup{colorlinks,citecolor= black,linkcolor= black,urlcolor=black}

\usepackage{amssymb,color,esvect,amsmath,graphicx,latexsym,amsthm,slashed,eso-pic,amsfonts,bbold}
\definecolor{darkgreen}{rgb}{0,0.6,0}
\definecolor{darkblue}{rgb}{0,0,0.5}
\usepackage[dvipsnames]{xcolor}
\usepackage{hyperref}

\usepackage{verbatim}
\definecolor{nicegreen}{rgb}{0.,0.5,0.}
\newcommand{\p}{\varphi}

\newcommand{\beq}{\begin{equation}}
\newcommand{\eeq}{\end{equation}}
\newcommand{\be}{\begin{equation}}
\newcommand{\ee}{\end{equation}}
\newcommand{\bea}{\begin{eqnarray}}
\newcommand{\eea}{\end{eqnarray}}
\newcommand{\bear}{\begin{eqnarray}}
\newcommand{\eear}{\end{eqnarray}}

\newcommand{\Ap}{A^\prime}
\newcommand{\m}{\text{ m}}
\newcommand{\cm}{\text{ cm}}
\newcommand{\mAp}{m_{A^\prime}}

\newcommand{\nl}{\nonumber \\}
\newcommand{\MeV}{\text{ MeV}}
\newcommand{\GeV}{\text{ GeV}}
\newcommand{\TeV}{\text{ TeV}}

\newcommand{\eV}{\text{ eV}}
\newcommand{\order}[1]{\mathcal{O}{(#1)}}
\newcommand{\ba}{\begin{array}}
\newcommand{\ea}{\end{array}}

\def\OMIT#1{}

\begin{document}

\def\Cincinnati{Department of Physics, University of Cincinnati, Cincinnati, Ohio 45221, USA}
\def\SLAC{SLAC National Accelerator Laboratory, Menlo Park, CA 94025, USA}

\preprint{SLAC-PUB-17238}


\title{Dark Sectors at the Fermilab SeaQuest Experiment}


\author{Asher Berlin}
\email[Electronic address:]{berlin@slac.stanford.edu}
\affiliation{\SLAC}

\author{Stefania Gori}
\email[Electronic address:]{stefania.gori@uc.edu}
\affiliation{\Cincinnati}

\author{Philip Schuster}
\email[Electronic address:]{schuster@slac.stanford.edu}
\affiliation{\SLAC}

\author{Natalia Toro}
\email[Electronic address:]{ntoro@slac.stanford.edu}
\affiliation{\SLAC}


\date{\today}

\begin{abstract}

We analyze the unique capability of the existing SeaQuest experiment at Fermilab to discover well-motivated dark sector physics by measuring displaced electron, photon, and hadron decay signals behind a compact shield. A planned installation of a refurbished electromagnetic calorimeter could provide powerful new sensitivity to GeV-scale vectors, dark Higgs bosons, scalars, axions, and inelastic and strongly interacting dark matter models. This sensitivity is both comparable and complementary to NA62, SHiP, and FASER. SeaQuest's ability to collect data now and over the next few years provides an especially exciting opportunity. 

\end{abstract}


\maketitle

\section{Introduction} 
\label{sec:intro}

Dark matter (DM) provides compelling evidence that we are overlooking new fundamental forms of matter and forces. 
While tremendous progress has been made looking for DM candidates
that are charged under known Standard Model (SM) forces (e.g.~WIMPs), searches to date have only provided powerful exclusions. 
This fact, as well as independent theoretical motivation, has triggered a great deal of effort
to search for DM candidates and associated interactions that are 
SM neutral -- dark sectors. Particular attention has been given to exploring new particle interactions at GeV-scales and below, 
and a vibrant set of efforts to look for dark photons, scalars, and a variety of DM models 
is now underway. Because dark sectors are expected to interact feebly with the SM, 
either through loop- or mass-suppressed operators, 
many models have long-lived particles that decay back to SM states with a detectable lifetime. 
In fact, many existing experiments aim to take advantage of such signatures, 
such as the Heavy Photon Search (HPS), various beam dump experiments, and searches at the LHC~\cite{Alexander:2016aln,Battaglieri:2017aum,ATLAS:2016jza,Aad:2015uaa,Khachatryan:2015vta,Aaij:2016xmb,Aaij:2016isa,Aaij:2016qsm,Aaij:2015tna,Celentano:2014wya}.

An especially fruitful direction of recent investigation has focused on the use of 
high-energy proton beams in fixed-target setups to produce and detect long-lived 
particles. Compared to electron beams,  proton beams offer several advantages. First, dark photon and dark Higgs production rates on a thick target are 
typically several orders of magnitude larger for proton beams than comparable-intensity electron beams, due to their higher penetrating power and enhancements from meson decay reactions and/or strong interactions. 
Second, proton reactions on a thick target also produce cascades of muons and mesons whose scattering in the beam dump serves as an additional signal production source 
for axions or leptophilic Higgs bosons.   
Third, there are several existing and planned high-energy proton beams that provide for a significant boost 
to enhance the laboratory lifetime of new states. 
To date, most work has focused on recasting results from past proton beam dump experiments~\cite{Blumlein:2013cua,Batell:2009di,Alexander:2016aln,Battaglieri:2017aum}, followed by 
a series of proposals for future experiments such as SHiP~\cite{Alekhin:2015byh} and FASER~\cite{Feng:2017uoz}. 
However, comparably little attention has been given to existing proton fixed-target experiments (aside from NA62~\cite{NA62:2017rwk}), especially those with 
access to meter-scale decay lengths.  

In this paper, we show that the existing SeaQuest experiment at the Fermi National Accelerator Laboratory (Fermilab), with only modest changes, can be used to search for particles with meter-scale lab-frame displacement,
with reach comparable and complementary to SHiP, FASER, and NA62.
The SeaQuest setup is rather unique for its compactness and its access to Fermilab's currently operational 120 GeV proton beam (see Sec.~\ref{sec:seaquest}). 
It is currently capable of detecting displaced decays into muons, and, in the near future, can be upgraded to detect electrons and identify charged pions. 
We discuss how ongoing and near term upgrades to SeaQuest can provide sensitivity to a wide variety of dark sectors, 
and we illustrate this by calculating sensitivity to several important benchmark models, including dark photons, dark Higgs bosons, inelastic DM, leptophilic scalars, and axion-like particles.  

\begin{figure*}[t] 
 \vspace{0.cm}
 \includegraphics[width=0.6\textwidth]{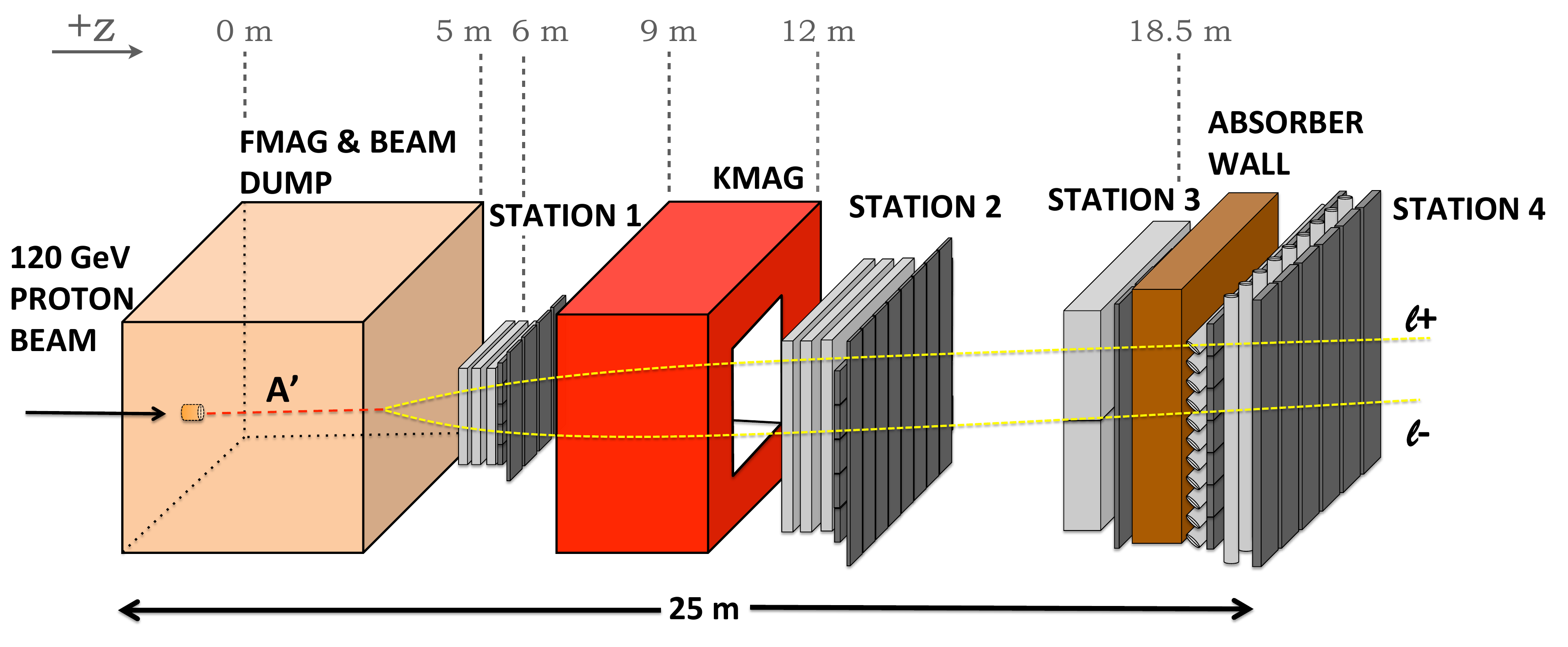}
\caption{Layout of the SeaQuest spectrometer in its current form (adapted from Ref.~\cite{Gardner:2015wea}).}
\label{fig:SeaQuestSetup}
\end{figure*}

The paper is organized as follows: In Sec.~\ref{sec:seaquest}, we introduce the SeaQuest experiment and discuss the possible upgrades of the present apparatus. In Sec.~\ref{sec:analysis}, we discuss general aspects of our analysis. Secs.~\ref{sec:darkphoton}--\ref{sec:iDM} focus on dark-photon-initiated signals.  In Sec.~\ref{sec:darkphoton}, we discuss the physics of dark photon production in proton fixed-target collisions and contrast with production in electron-target collisions. In Sec.~\ref{sec:SeaQuestAprime}, we analyze the prospects of detecting a minimal dark photon at SeaQuest (which has also been discussed in Ref.~\cite{Gardner:2015wea}). In Sec.~\ref{sec:iDM}, we study SeaQuest's sensitivity to dark photons decaying into inelastic DM. In Secs.~\ref{sec:leptoHiggs} and \ref{sec:other}, we briefly discuss the prospects for SeaQuest to detect models of leptophilic scalars, minimal dark Higgs bosons, and axion-like particles. Conclusions of this work are presented in Sec.~\ref{sec:conclusions}.

\section{The SeaQuest Experiment} \label{sec:seaquest}

The SeaQuest spectrometer is currently operating at Fermilab with access to the 120 GeV main injector proton beam~\cite{Aidala:2017ofy}. It is designed to study the sea quark content of the proton by measuring Drell-Yan dimuon production from the collision of protons with various nuclear and polarized targets. Recently, the experiment has seen the installation of a displaced vertex trigger~\cite{doi:10.1142/S0217732317300087,mingmaryland,Sanghoon}, allowing the detection of muons originating from the decays of exotic long-lived and low-mass particles. 

A schematic layout of the SeaQuest detector is shown in Fig.~\ref{fig:SeaQuestSetup}. The detector extends up to $\sim 25 \m$ in  length and is comprised of a series of tracking/triggering and muon-identification stations. A $5 \m$ long magnetized iron block (``FMAG") is placed $\lesssim 1 \m$ downstream from a thin nuclear target.\footnote{A 25 cm hole along the beam line is drilled into the front of FMAG, in order to spatially separate events originating from the nuclear target and the dump, without increasing single muon rates from the decay of charged pions in flight.} This serves as a focusing magnet and a beam dump for the relatively unattenuated proton beam. Its magnetic field imparts a kick of $\Delta p_T \simeq 2.9 \GeV$ and effectively sweeps away soft SM radiation, aside from, e.g., high-energy neutrinos, muons, and neutral mesons. An additional $3 \m$ long open-aperture magnet (``KMAG") is placed between the first two tracking stations and imparts a transverse momentum kick of $\Delta p_T \simeq 0.4 \GeV$ in order to facilitate accurate momentum reconstruction.

SeaQuest offers a unique combination of advantages compared to previous and existing high-intensity experiments. For instance, compared to electron beam dumps, SeaQuest benefits from large particle production rates. Compared to previous proton beam dumps, SeaQuest operates at a higher energy than LSND~\cite{Athanassopoulos:1996ds} ($\sim 120 \GeV$ vs. $\sim 0.8 \GeV$) and is sensitive to shorter decay lengths than CHARM~\cite{Allaby:1987vr} ($\sim 1 \m$ vs. $\sim 100 \m$). Other high-intensity proton beam experiments are expected to acquire data in the near and more distant future. For instance, NA62~\cite{NA62:2017rwk} and the proposed SHiP experiment at CERN~\cite{Alekhin:2015byh} will have access to the $400 \GeV$ SPS beam. However, these instruments will have a longer decay volume, thicker shielding, and a complementary sensitivity to longer lifetimes (see Table~\ref{Tab:ProtonFixedTarget} below). As we explore in this work, SeaQuest can potentially probe large regions of motivated and currently unexplored model space in the near future with minor upgrades to the existing spectrometer. 

A parasitic run at SeaQuest using the displaced vertex trigger recently acquired $\sim 3 \times 10^{16}$ protons on target (POT) of data in the search for long-lived particles~\cite{PrivateSeaQuest}. The signal is a muon pair that is significantly displaced from the front of FMAG. An additional run utilizing the displaced muon trigger is expected to begin at the end of 2018 and will acquire $\sim 1.44 \times 10^{18}$ POT in two years of parasitic data taking, equivalent to $\sim 35 \text{ ab}^{-1}$ of integrated luminosity~\cite{TalkStatusSeaQuest}. We will denote this luminosity phase as ``Phase I." As another benchmark luminosity, we also outline the SeaQuest reach with $10^{20}$ POT (``Phase II"), a dataset similar to that of MiniBooNE~\cite{Aguilar-Arevalo:2017mqx} and the proposed SHiP experiment, which could be collected in the coming years as a result of the Fermilab Proton Improvement Plan~\cite{Shiltsev:2017mle}.

At SeaQuest, there are plans to install a refurbished electromagnetic calorimeter (ECAL) from the PHENIX detector at Brookhaven National Laboratory within the next year~\cite{PrivateSeaQuest,doi:10.1142/S0217732317300087,mingmaryland}. This upgrade would allow SeaQuest to measure energetic electrons, enlarging the discovery potential for long-lived particles below the dimuon threshold. In this study, we discuss the physics goals that could be achieved after the proposed ECAL upgrade. The optimal location for the calorimeter within the spectrometer is uncertain, as is the specific form of the displaced electron trigger. For concreteness, we assume that the ECAL is installed between tracking stations 3 and 4, i.e., in place of the hadron absorber wall, as shown in Fig.~\ref{fig:SeaQuestSetup}. In this case, tracking in station 4 can be utilized for additional particle identification by functioning as a muon veto. As noted in Ref.~\cite{Gardner:2015wea}, it might be necessary to add an additional small magnet after FMAG in order to properly separate electron pairs. In the remainder of this work, we assume that the electrons are adequately separated and that SeaQuest's vertexing capability is efficient in this setup. 

An ECAL upgrade would also present an opportunity to conduct a search with lower (or at least very different) backgrounds than dimuon searches. A possible background in the muon channel consists of energetic pairs of muons that originate from prompt SM processes. These muons can multiple-scatter within the iron dump and emerge from the backend of FMAG as if they originated from a displaced vertex. The magnetic focusing effect of FMAG significantly reduces this background relative to naive expectations, but it has not yet been measured or computed in the low-mass phase-space and may be limiting. In contrast, prompt electrons are easily blocked in the thick iron dump so that fake displaced vertices never arise in the dielectron channel.  Hereafter, we will adopt the standard assumption taken by the SeaQuest collaboration, namely that electron pairs originating downstream of FMAG constitute a nearly background-free signal. We will discuss the validity of this assumption below. 

\section{General Analysis} 
\label{sec:analysis}

In this section, we discuss our assumptions regarding SeaQuest's sensitivity to displaced vertices and our modeling of the experimental acceptance.  In Sec.~\ref{ssec:fiducial}, we describe three possible fiducial decay regions that may allow for low-background dielectron searches at SeaQuest.  We discuss backgrounds from true displaced vertices (e.g.~semileptonic $K^0_L$ decays), which are argued to be negligible for the tightest decay region but may be a concern for the looser fiducial regions.  Nonetheless, we will assume throughout this work that backgrounds can be reduced to the $\order{1}$-event level.  Therefore, our sensitivity projections for SeaQuest correspond to regions of parameter space where at least 10 signal events are expected.  In Sec.~\ref{ssec:acceptance}, we discuss our treatment of the geometric detector acceptance and the effects of KMAG on particle trajectories.  We emphasize that the geometric acceptance and decay length distributions cannot be factorized because the acceptance favors highly boosted tracks that are less strongly deflected by KMAG.

\subsection{Fiducial Decay Regions and Background Considerations}
\label{ssec:fiducial}

We will investigate three possible fiducial decay regions (as measured from the upstream end of FMAG): 
\begin{itemize}
\item \underline{$5 \m - 6 \m$}: After FMAG and before station 1. 
\item \underline{$5 \m - 9 \m$}: After FMAG and before KMAG. 
\item \underline{$5 \m - 12 \m$}: After FMAG and before the end of KMAG. In this case, KMAG only exerts a partial $\Delta p_T$ kick, corresponding to the fraction of the magnet traversed by the electron pair.
\end{itemize}
Throughout this work, we will investigate the physics implications for searches in each of these setups assuming negligible background, though we stress that this might be too optimistic for the larger decay regions. 

Electrons originating in the $5 \m - 6 \m$ region bend in the magnetic field of KMAG, which allows for accurate momentum and pointing measurements after tracking in stations 1-3. This is the most conservative fiducial region.

In the $5 \m - 9 \m$ decay region, electrons need not pass through station 1, and, hence, momentum reconstruction is not possible with KMAG alone. In this case, energy deposition in the ECAL  provides the only estimate of the electrons' energies, degrading the ability to accurately reconstruct the location of the primary vertex and momentum of its progenitor.  However, if backgrounds can nonetheless be kept to sufficiently low levels, the larger decay volume is advantageous.

This is also the case for the $5 \m - 12 \m$ decay region.  In addition, decay products emerging \emph{within} KMAG receive a smaller $\Delta p_T$ kick.  Therefore, in parts of this decay region even very soft final-state particles can enter the detector acceptance.  For example, a forward-going electron must have energy $E_e \gtrsim 2 \GeV$ to fall within the detector acceptance if it is produced immediately upstream of KMAG; if it only experiences $1/4$ of the KMAG field, this energy threshold drops to $\sim 0.5 \GeV$.  This has minor impact on the acceptance of visibly decaying dark photons or dark Higgs bosons, but significantly enhances the acceptance of lower-energy signals like the inelastic DM models considered in Sec.~\ref{sec:iDM}.  Of course, it also allows for contributions from low-energy backgrounds that are not relevant for the other two fiducial decay regions.  Studies of the $5 \m - 12 \m$ region also give some indication of how the yield-limited sensitivity of SeaQuest --- even with a smaller decay region --- would change if the KMAG magnetic field was reduced.

A potentially important background to the displaced electron search emerges from semileptonic decays of neutral long-lived kaons~\cite{Anelli:2015pba}. 
An energetic kaon can penetrate FMAG before decaying ($K_L^0 \to \pi^\pm e^\mp \nu$) in one of the fiducial regions. Such events are signal-like if the charged pion is misidentified as an electron in the ECAL. A simple estimate shows that this process is not of concern in a search for displaced electrons in the $5 \m - 6\m$ fiducial region and illustrates the potential for greater backgrounds in the looser decay regions.  Approximately $10^{17}$ kaons are produced from $10^{18}$ POT. The pion interaction length in iron is $\sim 20 \cm$, and, hence, $\sim 10^6$ kaons are expected to traverse the entirety of FMAG. If $10 \%$ of these kaons decay within $5 \m  - 6 \m$ and $\sim 1 \%$ of the decay products are sufficiently energetic to remain in the geometric acceptance of the spectrometer, a pion rejection factor of $\lesssim 1 \%$ is needed to reduce this background to $\lesssim 10$ events. Fortunately, this level of sensitivity is feasible after the planned ECAL upgrade~\cite{Aphecetche:2003zr}. Additional pointing cuts can reduce this background to negligible levels. 

For the larger decay region of $5 \m - 9 \m$, we expect that the reduction in the pointing resolution and the increased likelihood for the kaon to decay within the signal region can potentially lead to enhanced background rates (as large as a few events if pointing is completely ineffective). Additionally, in the $5 \m - 12 \m$ region, the increased geometric acceptance of soft kaon decays could lead to substantially larger backgrounds that can only be rejected by pointing (limited by the ECAL energy resolution).  Therefore, assuming a sensitivity to 10 signal events for the $5 \m - 12 \m$ region is likely unrealistic.  Nonetheless, it is useful as a proxy for minor detector modifications that could be possible in dedicated runs, which maintain low-energy signal acceptance while mitigating the kaon-decay background.   For example, adding $\sim 1 \m$ of additional iron upstream of FMAG would decrease the total kaon yield by two orders of magnitude with little impact on the signal acceptance.  Alternately, using a smaller decay region upstream of station 1 (to improve pointing resolution) while reducing the KMAG field offers another approach to rejecting soft kaon decays.  Although it is beyond the scope of this work, these considerations motivate a more dedicated detector study in order to fully optimize the signal decay region and magnetic field strength of KMAG, while maintaining the ability to successfully discern signal from background. 

\subsection{Modeling of Geometric Acceptance and its Decay-Dependence}
\label{ssec:acceptance}

We demand that a long-lived particle decays to electrons in one of the three fiducial regions outlined above and that the displaced electrons traverse the entire remaining geometry of the detector after bending through the KMAG magnet. Since the geometric acceptance depends on the particular location of the decay, the total efficiency is given by
\be
\label{eq:TotalEff}
\text{effic.} = m \, \Gamma \, \int_{z_{\text{min}}}^{z_{\text{max}}}d z \, \sum_{\text{events $\in$ geom.}} \frac{e^{-z \, (m  / p_z) \, \Gamma}}{N_\text{MC} \, p_z} 
~,
\ee
where $z_\text{min} - z_\text{max}$ defines the fiducial decay region along the $\hat{z}$ (beam) direction and $z$ is the position of the decay. $z_\text{min} = 5 \m$ and $z_\text{max}=6 \m, \, 9 \m, \, 12 \m$ for the three fiducial regions outlined above. $\Gamma$, $m$, and $p_z$ are the width, mass, and the $\hat{z}-$component of the momentum of the long-lived decaying particle. $N_\text{MC}$ denotes the total number of simulated events in a Monte Carlo sample, and the sum is performed over only the subset of those events that pass the $z$-dependent geometric cut. We include events in which both electrons are captured by tracking station 3, which is located $\sim 18.5$ m downstream of the target side of FMAG and approximated by a $\sim 2\m \times 2 \m$ square in the transverse plane. KMAG is modeled as an instantaneous transverse kick, $\Delta p_T \simeq 0.4 \GeV \times (\Delta z_K / 3 \m)$ along the $\hat{x}$ direction, where $\Delta z_K$ is the distance traversed by the electron pair through the magnet. We have checked that this selection criteria is in close agreement with those adopted by the SeaQuest collaboration in their preliminary simulations~\cite{PrivateSeaQuest}. 

The expected number of signal events is given by
\be
\label{eq:Nsignal}
N_\text{signal} = N \times \text{BR} (e^+ e^-) \times \text{effic.}
~,
\ee
where $N$ is the total number of long-lived particles produced and $\text{BR} (e^+ e^-)$ is its branching ratio to electrons. We will utilize Eqs.~(\ref{eq:TotalEff}) and (\ref{eq:Nsignal}) throughout our analysis. In the specific limit that the geometric criteria is independent of the decay position and $p_z$, Eq.~(\ref{eq:TotalEff}) reduces to the factorized form that is often quoted in the literature,
\be
\label{eq:TotalEff2}
\text{effic.} \xrightarrow{\text{factorize}} \mathcal{A} \, \left\langle e^{-z_\text{min}  ( m / p_z )  \Gamma} - e^{-z_\text{max}  ( m/ p_z )  \Gamma}  \right\rangle
,
\ee
where $\mathcal{A}$ and the brackets denote the geometric acceptance and an average over all simulated events, respectively. However, if, e.g., the geometric cuts are significantly (anti-)correlated with the boost or decay position, Eq.~(\ref{eq:TotalEff2}) is not an effective approximation of the more general form in Eq.~(\ref{eq:TotalEff}). At SeaQuest, demanding that the electrons successfully punch through the magnetic field of KMAG results in a strong preference for highly boosted events. As a result, the geometric efficiency and position of the decay are strongly correlated, and the limiting form in Eq.~(\ref{eq:TotalEff2}) is not an accurate approximation of Eq.~(\ref{eq:TotalEff}).  
Indeed, calculations utilizing Eq.~(\ref{eq:TotalEff2}) underestimate the projected reach of SeaQuest for long-lived states whose lifetime is short compared to the fiducial baseline. This will be discussed in more detail in Sec.~\ref{sec:SeaQuestAprime}.

\section{Dark Photon production at SeaQuest} 
\label{sec:darkphoton}

A well-motivated new force carrier is the hypothetical dark photon. 
In the minimal model, a new broken $U(1)_D$ symmetry is added to the SM. The corresponding gauge boson, denoted as $\Ap$, couples to SM hypercharge through the kinetic mixing term~\cite{Okun:1982xi,Holdom:1985ag}
\beq\label{eq:KinMix}
\mathcal{L} \supset \frac{\epsilon}{2 \, \cos{\theta_w}} ~ A^{\prime}_{\mu \nu} ~ B^{\mu \nu}
~,
\eeq
where $\theta_w$ is the Weinberg angle.
We remain agnostic about the mechanism that generates the dark photon mass, $\mAp$. 
Although $\epsilon$ is a free parameter of the low-energy theory, it is natural to expect loop-induced mixings ($\epsilon \lesssim 10^{-3}$) if there exist any particles charged under both $U(1)_\text{Y}$ and $U(1)_D$~\cite{Holdom:1985ag,delAguila:1988jz}. 
In the $\mAp \lesssim \GeV$ mass range of interest for SeaQuest (which can arise, for example, from a quartic coupling of a $U(1)_D$-charged scalar with the SM Higgs as in supersymmetric embeddings~\cite{Cheung:2009qd}),  $\mAp \ll m_Z$ implies that $\Ap$ dominantly mixes with the SM photon below the scale of electroweak symmetry breaking. As a result, SM fermions of electric charge $e \, Q_f$ inherit a millicharge under $U(1)_D$, $\epsilon \, e \, Q_f$. From a bottom-up perspective, this minimal model is completely governed by the two free parameters $\mAp$ and $\epsilon$, and constraints can be placed in the $\mAp-\epsilon$ plane. 

At SeaQuest, most dark photons are produced in primary reactions resulting from the collision of the proton beam with the iron beam dump. Within a single nuclear collision length of iron ($X_c \simeq 82 \text{ g} / \text{cm}^2$), the effective luminosity for proton-proton collisions is given by
\be
\label{eq:lum}
L \simeq \frac{Z \, (X_c / \text{g}) \, N_A}{A} \, \text{POT} \simeq 35 \text{ ab}^{-1} \left( \frac{\text{POT}}{1.44 \times 10^{18}} \right)
,
\ee
where $A (Z) = 56 (26)$ is the atomic mass (number) of iron and $N_A$ is Avogadro's number.

\begin{figure}[t] 
 \vspace{0.cm}
 \includegraphics[width=0.5\textwidth]{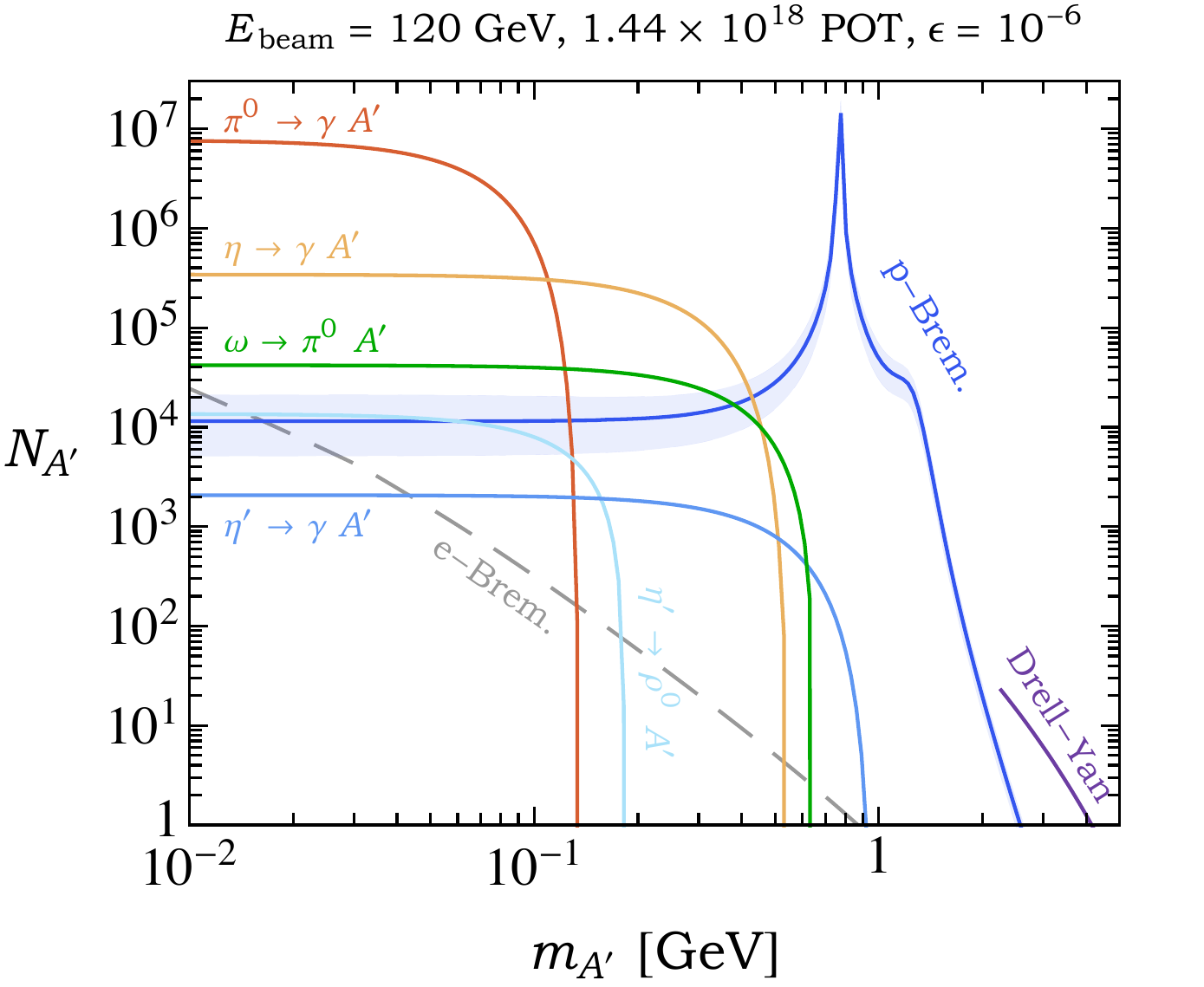}
\caption{Number of dark photons (solid color) produced at Phase I of SeaQuest ($1.44\times 10^{18}$ POT) in various production channels for $\epsilon=10^{-6}$. An estimate of the theory uncertainty for proton Bremsstrahlung is shown as the shaded blue region (see text for details). For comparison, we also show the analogous production rate for electron Bremsstrahlung (dashed gray), assuming a 120 GeV electron beam, $1.44 \times 10^{18} \text{ EOT}$, and production within the first radiation length of a tungsten target.
}
\label{figure:productionAprime}
\end{figure}

Proton fixed-target experiments benefit from large rates compared to electron beams.  This is illustrated in Fig.~\ref{figure:productionAprime}, which shows, as a function of dark photon mass, the contributions of different meson decays, Bremsstrahlung, and Drell-Yan processes to dark photon yields (normalized to SeaQuest's Phase I luminosity and $\epsilon=10^{-6}$).  For all dark photon masses, the yield per  (high-energy) proton incident on a thick target is orders of magnitude larger than the yield per electron.  This enhancement is attributable to several factors: (1) Protons are more penetrating than electrons, and so interact with a larger number of target nuclei.  (2) In the GeV dark photon mass range, proton-initiated Bremsstrahlung is enhanced relative to electron-initiated Bremsstrahlung by the ratio of couplings $\alpha_s/\alpha_\text{em}$. (3) Finally, light mesons are produced in large numbers, and their branching fractions to lower-mass dark photons are suppressed only by $\epsilon^2$ and not also by $\alpha_\text{em}$~\cite{Reece:2009un}.

Both the yields and the kinematics of these various production modes will be important in the following studies.  We now discuss these production modes in detail. 

\subsection{Meson Decays}
Due to the substantial energy of the Fermilab beam, the rate for meson production is large, 
even for relatively heavy mesons, such as the $\eta^\prime$. In the SM, several mesons have a sizable branching fraction into at least one photon. This gives rise to $\Ap$ production through kinetic mixing with the SM photon. Examples of such decays involving SM photons include
\begin{align}\label{eq:processes}
& \pi^0 \, ,\, \eta \, ,  \, \eta^\prime \to\gamma\gamma~,
\nl
& \omega\to\pi^0\gamma~,
\nl
& \eta^\prime \to \rho^0 \gamma 
~.
\end{align}
We model the production of SM mesons in {\tt PYTHIA 8.2}~\cite{Sjostrand:2014zea}. In order to validate this simulation, we compare our rates and spectra from {\tt PYTHIA}  
 with the observed meson production at high-energy proton beams, such as the 120 GeV beam of the Main Injector Particle Production (MIPP) experiment at Fermilab~\cite{Mahajan:2013awa} and the 400 GeV SPS proton beam at CERN~\cite{Bonesini:2001iz}. Our simulated rates for $\pi^0$ and $\eta$ production are within $\sim 50\%$ of the measured values. For heavier mesons ($\phi$ and $\Omega$), the agreement is less accurate but still within a factor of $\sim 2$. This ambiguity 
translates into a small uncertainty for our projected dark photon sensitivity. For a 120 GeV beam, we obtain production rates of $N_\pi\sim 3.5$, $N_\eta\sim 0.4$, $N_{\eta^\prime}\sim 0.04$, and $N_\omega\sim 0.45$ per proton-proton collision. We have also checked that the kinematic spectra of meson production are in relatively good agreement with the measured distributions. 
The relative importance of each decay can be characterized by the product of the number of mesons produced per proton and the corresponding branching fraction into SM photons, e.g.,
\begin{align}
& N_\pi  \times \text{BR} ( \pi^0 \to \gamma \gamma ) \sim 3.5~,
\nl
& N_{\eta}  \times \text{BR} ( \eta \to \gamma \gamma ) \sim 0.15~,
\nl
& N_{\omega}  \times \text{BR} ( \omega \to \pi^0 \gamma ) \sim 5 \times 10^{-2}~,
\nl
& N_{\eta^\prime}  \times \text{BR} ( \eta^\prime \to \rho^0 \gamma ) \sim 10^{-2}~,
\nl
& N_{\eta^\prime}  \times \text{BR} ( \eta^\prime \to \gamma \gamma ) \sim 10^{-3}
~.
\end{align}

Analogous processes involving dark photons are obtained by the substitution $\gamma \to \Ap$. From the meson spectra generated in  {\tt PYTHIA}, we manually decay the mesons to final states involving an $\Ap$. Processes involving two dark photons in the final state are suppressed by additional powers of $\epsilon$. The relevant meson branching fractions are rescaled from the SM values by $\epsilon^2$ and phase-space factors~\cite{deNiverville:2011it,Gardner:2015wea}. The branching ratios are given by
\begin{widetext}
\begin{align}
\label{eq:pions}
&\text{BR}(\pi^0 , \eta, \eta^\prime \to \gamma \Ap) \simeq 2 \, \epsilon^2 \, \Big( 1 - \frac{\mAp^2}{m_{\pi^0, \eta, \eta^\prime}^2}\Big)^3
\times \text{BR}(\pi^0, \eta, \eta^\prime \to \gamma \gamma)
\nl
&\text{BR}(\eta^\prime \to \rho^0 \Ap) \simeq \epsilon^2 \, \left( m_{\eta^\prime}^2 - m_\rho^2\right)^{-3} 
\Big[ (\mAp^2 - (m_{\eta^\prime} + m_\rho)^{2}) (\mAp^2 - (m_{\eta^\prime} - m_\rho)^2) \Big]^{3/2}
\times \text{BR}(\eta^\prime \to \rho^0 \gamma)
\nl
&\text{BR}(\omega \to \pi^0 \Ap) \simeq \epsilon^2 \, \left( m_{\omega}^2 - m_\pi^2\right)^{-3} \Big[ (\mAp^2 - (m_\pi + m_\omega)^{2}) (\mAp^2 - (m_\pi - m_\omega)^2) \Big]^{3/2} \times \text{BR}(\omega \to \pi^0 \gamma)
~.
\end{align}
\end{widetext}
We have checked that other processes such as $\eta^\prime \to \omega \Ap$ and $\phi \to \eta \Ap$ are subdominant to the ones considered above. 

The number of dark photons produced for different values of $\epsilon$ and the accumulated luminosity can be simply rescaled from Fig.~\ref{figure:productionAprime} as $N_{\Ap} \propto \epsilon^2 \times {\text{POT}}$. %
For $\mAp \lesssim 100 \MeV$, $\Ap$ production is dominated by exotic $\pi^0$ decays, with
\be
\label{eq:NApPion}
N_{\Ap}(\text{pion decay}) \sim 10^{7} \times \left( \frac{\epsilon}{10^{-6}} \right)^2 \left( \frac{\text{POT}}{10^{18}} \right)
~.
\ee
The decay of $\eta$ mesons has roughly $1/20$ this yield, but dominates for $100 \MeV \lesssim \mAp \lesssim 500 \MeV$. 
The lower yields and photon branching ratios of heavier mesons make their contributions to dark photon production subdominant.  In addition, many of these processes for heavier mesons involve final state mesons, so that the accessible dark photon mass range is limited by \emph{mass differences}.  For example, $\phi \to \eta \, \Ap$ is only possible for $m_{\Ap}<m_\phi - m_\eta$ and is therefore relevant in a smaller region of phase-space than $\eta \to \gamma \, \Ap$ decays.

\subsection{Bremsstrahlung}
As shown in Fig.~\ref{figure:productionAprime}, the majority of dark photons are produced through proton Bremsstrahlung for $\mAp \gtrsim 500 \MeV$. 
In calculating Bremsstrahlung production, we follow the procedure outlined in Refs.~\cite{deNiverville:2016rqh,Gorbunov:2014wqa,Blumlein:2013cua}, manually generating events that are weighted by the relevant differential cross-section, $d^2 \sigma/(dz~dp_T^2)$, where $p_T$ and $z$ are the transverse momentum and the fraction of the beam momentum carried by the outgoing dark photon, respectively. As discussed in Refs.~\cite{deNiverville:2016rqh,Gorbunov:2014wqa,Blumlein:2013cua}, this analysis is valid as long as $E_\text{beam},~E_{A^\prime},~E_\text{beam}-E_{A^\prime}\gg m_p,~|p_T|$, where $E_\text{beam}$ and $E_{A^\prime}$ are the proton beam and dark photon energy, respectively. 
These kinematic conditions lead to a restricted range for $z$, as well as an upper bound on $p_T$. For a 120 GeV proton beam, these conditions are $z\in ( \sim 0.1, \sim 0.9)$ and $p_T\lesssim 1$ GeV. We have checked that our results do not depend significantly on the precise values of these limits, 
aside from the minimum value of $z$. For this reason, we  
choose a generous range for this lower bound ($z\geq0.2$ and $z\geq0.05$), leading to the blue shaded band in Fig.~\ref{figure:productionAprime}. This serves as an estimate for the theoretical uncertainty in our Bremsstrahlung calculation.

For $\mAp \lesssim \text{GeV}$, the Bremsstrahlung cross-section is parametrically of size
\be
\sigma \sim \alpha_\text{em} \, \epsilon^2 \times \sigma_{pp} 
~,
\ee
where $\sigma_{pp} \sim 50 \text{ mb}$ is the inelastic proton-proton scattering cross-section~\cite{Blumlein:2013cua,Patrignani:2016xqp}. We thus expect
\be
\label{eq:NApBrem}
N_{\Ap}(\text{$p$ Brem.}) \sim 10^{4} \times \left( \frac{\epsilon}{10^{-6}} \right)^2  \left( \frac{\text{POT}}{10^{18}} \right).
\ee
Even for values of $\epsilon$ as small as $10^{-6}$, at least $\order{10^4}$ dark photons are produced from proton Bremsstrahlung in the mass range $0.5 \GeV \lesssim \mAp\lesssim 1.5 \GeV$. In Fig.~\ref{figure:productionAprime}, the enhancement at $\mAp \sim 800 \MeV$ arises from resonant mixing between the dark photon and SM $\rho^0$~\cite{deNiverville:2016rqh}. 

For comparison, at electron fixed-target experiments, the cross-section for $\Ap$ production via electron Bremsstrahlung takes the parametric form~\cite{Bjorken:2009mm}
\be
\sigma \sim \frac{\alpha_\text{em}^3 \, \epsilon^2}{\mAp^2} ~Z^2 
~,
\ee
where we have ignored various logarithmic and $\order{1}$ factors that modify the expression above. For $\Ap$ production within one radiation length of tungsten and a comparable number of electrons on target (EOT), we find
\begin{align}
N_{\Ap}(\text{$e$ Brem.}) &\sim \left( \frac{\epsilon}{10^{-6}} \right)^2 \left( \frac{\mAp}{\text{GeV}} \right)^{-2} \left( \frac{\text{EOT}}{10^{18}} \right)
~.
\end{align}
If $\mAp \sim 1 \MeV - 100 \MeV$, this rate is significantly smaller than that of pion decay in Eq.~(\ref{eq:NApPion}). For $\mAp \gtrsim 100 \MeV$, dark photon production from pion decays is kinematically suppressed. However, for these masses the rate for electron Bremsstrahlung is much less than that of proton Bremsstrahlung in Eq.~(\ref{eq:NApBrem}). Hence, one generically expects a much larger dark photon production rate at proton beam dumps than at electron fixed-target experiments.

\subsection{Drell-Yan}
Drell-Yan production of dark photons is potentially significant for $\mAp \gtrsim \text{few} \times \text{GeV}$. In modeling this process, we have generated events with {\tt MadGraph5}~\cite{Alwall:2011uj} using the {\tt FeynRules}~\cite{Alloul:2013bka} model of Ref.~\cite{DarkPhotonMadGraph}. 
For $\mAp \lesssim \sqrt{s}  \simeq 15 \GeV$, many dark photons could be produced at Phase I and II of the SeaQuest experiment. Parametrizing $q \bar{q} \to \Ap$ as
\be
\sigma \sim \frac{\alpha_\text{em} \, \epsilon^2}{\mAp^2}
~,
\ee
we find
\be
N_{\Ap} (\text{Drell-Yan}) \sim \left( \frac{\epsilon}{10^{-6}} \right)^2 \left( \frac{\mAp}{\text{GeV}} \right)^{-2} \left( \frac{\text{POT}}{10^{18}} \right)
~.
\ee

\subsection{Estimating Minimal Dark Photon Sensitivity }

\renewcommand{\arraystretch}{1.5}
\begin{table}
\centering
\begin{tabular}{|c||c|c|c|c|c|}
\hline
 & $E_\text{beam}$ & $p_\text{min} $& POT & $z_\text{min}$ & $z_\text{max}$\\
\hline\hline
SeaQuest & 120 GeV & 10 \GeV & $10^{18}- 10^{20}$& $5 \m$ & $10 \m$\\ \hline
NA62 & 400 GeV & - & $10^{18}$ & $100 \m$& $250 \m$ \\ \hline
SHiP & 400 GeV & 100 \GeV & $10^{20}$ & $65 \m$& $125 \m$ \\ \hline
FASER & 6500 GeV & 1 \TeV & $10^{16} - 10^{17} $ & $390 \m$& $400 \m$ \\ \hline
\end{tabular}
\caption{A summary of relevant features of upcoming high-intensity proton fixed-target experiments. Using Eqs.~(\ref{eq:epsestimate1}) and (\ref{eq:epsestimate2}), these parameters can be used to estimate the experimental sensitivity to decays of long-lived particles, such as dark photons.}
\label{Tab:ProtonFixedTarget}
\end{table}

We conclude this section with a brief estimate of various proton fixed-target experiments' sensitivity to the decays of long-lived dark photons into SM leptons, $\Ap \to \ell^+ \ell^-$. This will be discussed in more detail in Sec.~\ref{sec:SeaQuestAprime}, and the rate for this decay is shown below in Eq.~(\ref{eq:ApDecaytoLeptons}). For production of dark photons via pion decay (see Eq.~(\ref{eq:NApPion})), the range of couplings that an experiment is sensitive to (for $\mAp \lesssim 100 \MeV$) is approximately given by
\be
\epsilon_\text{min} \lesssim \epsilon \lesssim \epsilon_\text{max}
~,
\ee
where
\be
\label{eq:epsestimate1}
\epsilon_\text{min} \sim \left( \frac{p_\text{min}}{\alpha_\text{em} \, \mAp^2 \, z_\text{max} \, \text{POT}} \right)^{1/4} 
~,
\ee
and
\be
\label{eq:epsestimate2}
\epsilon_\text{max} \sim \left( \frac{E_\text{beam} \, \log{\left[\frac{E_\text{beam} \, \text{POT}}{\alpha_\text{em} \, \mAp^2 \, z_\text{min}}\right]}}{\alpha_\text{em} \, \mAp^2 \, z_\text{min}} \right)^{1/2}
~.
\ee
Above, $z_\text{min} - z_\text{max}$ defines the fiducial decay region of a given experiment, $E_\text{beam}$ the energy of its proton beam, and $p_\text{min}$ the minimum momentum of the $\Ap$ that is needed for its decay products to be efficiently detected. Characteristic values of these inputs are shown in Table~\ref{Tab:ProtonFixedTarget} for SeaQuest as well as NA62~\cite{na62cern} and the proposed SHiP~\cite{Alekhin:2015byh} and FASER~\cite{Feng:2017uoz} experiments. For Phase I of SeaQuest, this parametric estimate roughly gives
\begin{align}
\epsilon_\text{min}^{\rm{SeaQuest}} &\sim  10^{-7} \times (\mAp / 10 \MeV)^{-1/2} 
\nl
\epsilon_\text{max}^{\rm{SeaQuest}} &\sim  10^{-4} \times (\mAp / 10 \MeV)^{-1}
~,
\end{align}
while for the high-luminosity runs of SHiP and FASER, we find
\begin{align}
\epsilon_\text{min}^{\rm{SHiP}}&\sim  10^{-8} \times (\mAp / 10 \MeV)^{-1/2} 
\nl
\epsilon_\text{max}^{\rm{SHiP}} &\sim  10^{-4} \times (\mAp / 10 \MeV)^{-1}
~,
\end{align}
and
\begin{align}
\epsilon_\text{min}^{\rm{FASER}}&\sim  10^{-6} \times (\mAp / 10 \MeV)^{-1/2} 
\nl
\epsilon_\text{max}^{\rm{FASER}}&\sim  10^{-4} \times (\mAp / 10 \MeV)^{-1}
~,
\end{align}
respectively. We have refrained from making a similar estimate for NA62 because this requires more detailed knowledge of their detector efficiency. As we will show in Figs.~\ref{fig:visAp} and \ref{fig:visApFuture}, these estimates are in rough agreement with detailed numerical calculations.

\begin{figure}[t] 
 \vspace{0.cm}
 \includegraphics[width=0.5\textwidth]{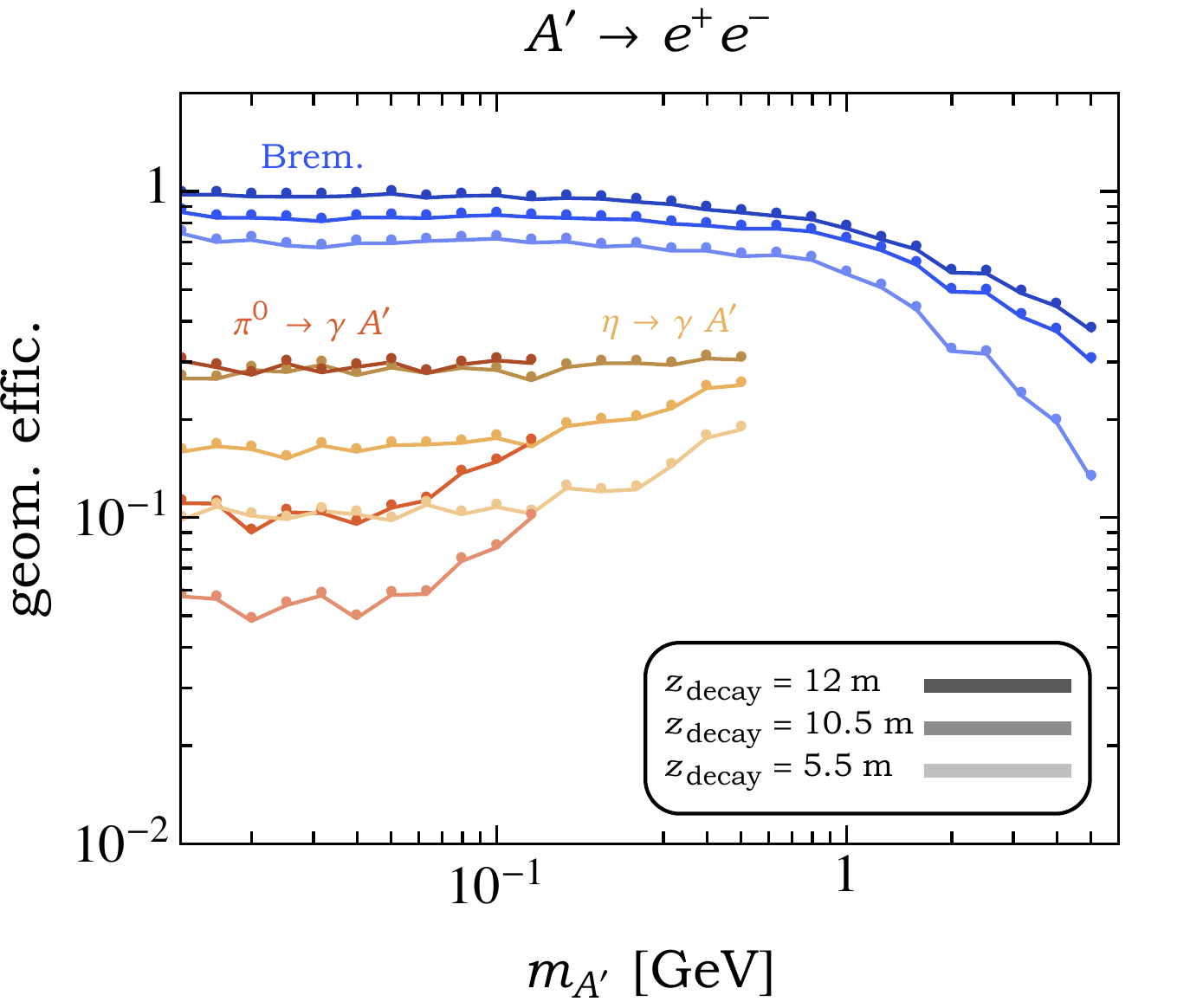}
\caption{In minimal dark photon models, the geometric efficiency for dark photons produced from proton Bremsstrahlung (blue) and the decays of pions (red) or eta mesons (orange). For each of these production channels, we assume that the $\Ap$ decays to an electron pair after traveling $5.5 \m$ (bottom line), $10.5 \m$ (middle line), or $12 \m$ (top line). For these latter two decay points, the electron pair only traverses a fraction of the KMAG magnet.}
\label{fig:efficiencyDarkPhoton}
\end{figure}
\begin{figure*}[t] 
\vspace{0.cm}
\includegraphics[width=0.497\textwidth]{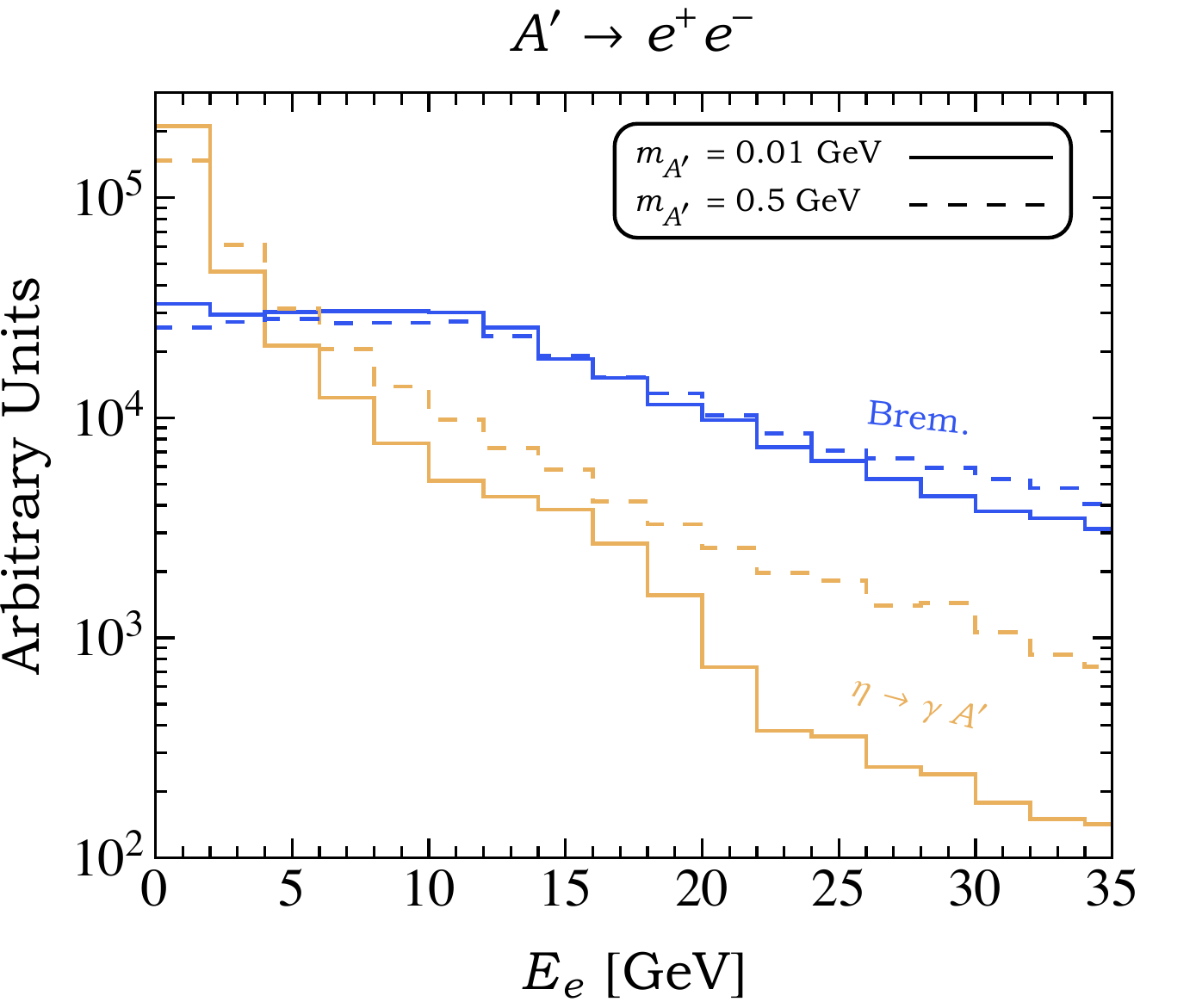}
\includegraphics[width=0.497\textwidth]{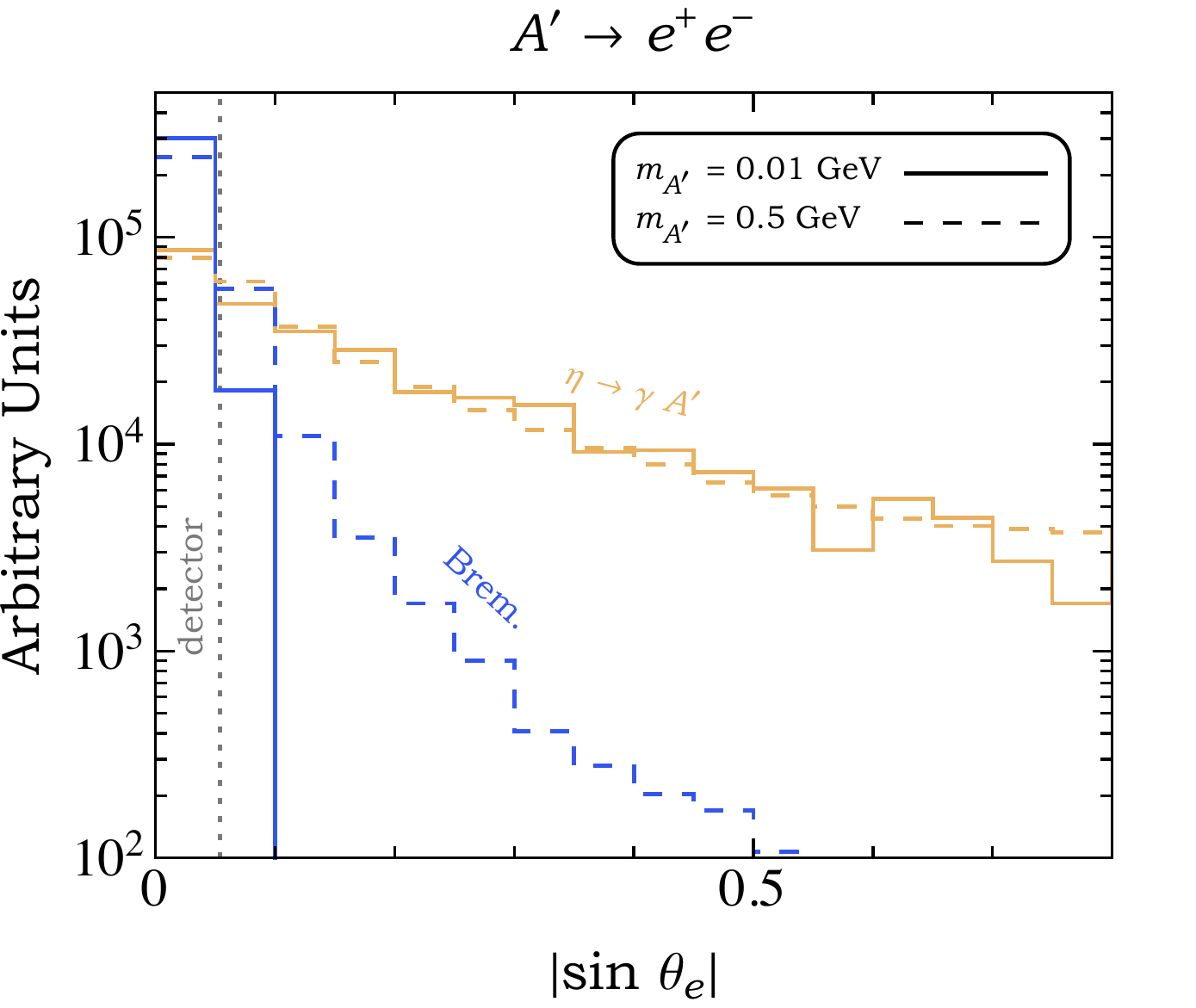}
\caption{Signal kinematics of $\Ap \to e^+ e^-$ for dark photons produced from exotic eta meson decays (orange) and proton Bremsstrahlung (blue). The left (right) panel displays energy (angular) distributions for electrons originating from dark photon decays before traveling through KMAG. The solid (dashed) line corresponds to $\mAp = 0.01 \GeV$ $(0.5 \GeV)$. The vertical gray dotted line in the right panel denotes the angular scale of the SeaQuest spectrometer.}
\label{fig:KinDarkPhoton}
\end{figure*}
%

\section{Minimal Dark Photon} 
\label{sec:SeaQuestAprime}

In the minimal dark photon model, the $A^\prime$ is the lightest state of the dark sector and can only decay to SM particles. Such decays are controlled by kinetic mixing, as in Eq.~(\ref{eq:KinMix}). For $m_\ell \ll \mAp \ll m_Z$, the partial width for decays to a pair of SM leptons ($\ell$) is approximately
\be
\label{eq:ApDecaytoLeptons}
\Gamma (\Ap \to \ell^+ \ell^-) \simeq \frac{ \alpha_\text{em} \, \epsilon^2}{3} ~ \mAp 
~.
\ee
For decays to hadronic final states, we take 
\be
\label{eq:ApDecaytoHadrons}
\frac{\Gamma (\Ap \to \text{hadrons})}{\Gamma (\Ap \to \mu^+ \mu^-)} \simeq R(\sqrt{s} = \mAp) 
~,
\ee
where $R \equiv \sigma (e^+ e^- \to \text{hadrons}) / \sigma(e^+ e^- \to \mu^+ \mu^-)$ is the data-driven parameter from Ref.~\cite{Olive:2016xmw}. For most values of $\mAp$ that we consider in this section, $\text{BR} (\Ap \to e^+ e^-) \gtrsim 10 \%$. Furthermore, the $\Ap$ proper lifetime is macroscopic ($\tau_{\Ap} \sim \text{cm}$) for $\epsilon \sim 10^{-6} \times (\mAp / \text{GeV})^{-1/2}$, and displaced leptonic decays can be efficiently searched for at SeaQuest.

As discussed in Sec.~\ref{sec:analysis}, we require that after traveling at least $5 \m$ from the target (upstream end of FMAG), the $\Ap$ decays to an electron pair that remains within the geometry of the spectrometer. The efficiency and total number of signal events are calculated using Eqs.~(\ref{eq:TotalEff}) and (\ref{eq:Nsignal}). 
In Fig. \ref{fig:efficiencyDarkPhoton}, the fraction of signal events that pass the geometric selection is presented for the dominant production channels as a function of $\mAp$. The position at which $\Ap$ decays to an electron pair is fixed to the representative lengths: $5.5 \m$ (bottom line), $10.5 \m$ (middle line), and $12 \m$ (top line). The geometric acceptance is enhanced for these latter two decay positions for two reasons: The larger solid angle subtended by the downstream tracking (in particular, station 3) for late decays increases angular acceptance.  Moreover, particles produced in or downstream of KMAG traverse a lower magnetic field integral, and their transverse deflection is correspondingly reduced, increasing the spectrometer's energy acceptance.

\begin{figure*}[t]
\hspace{-0.5cm}
\includegraphics[width=0.52\textwidth]{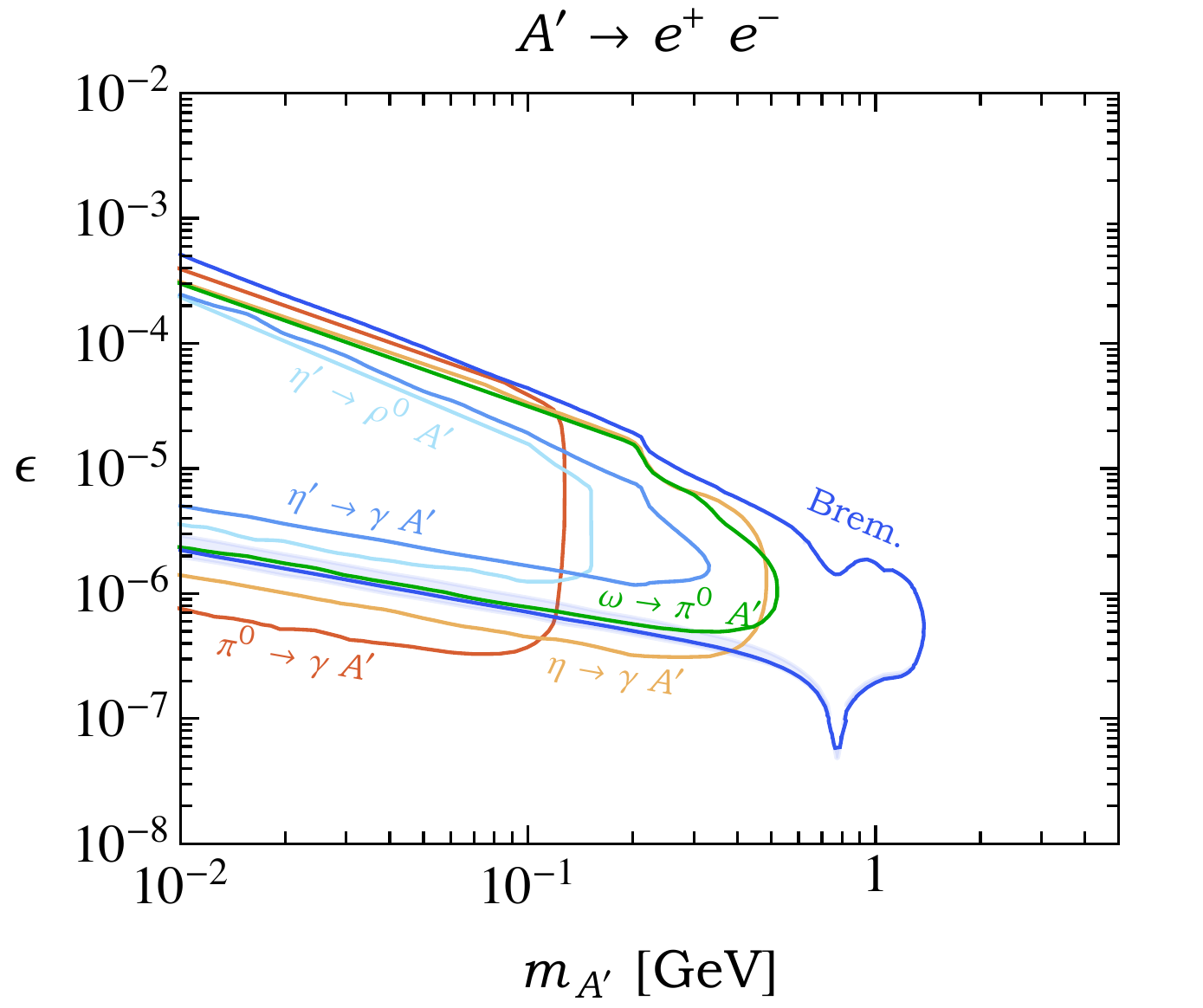} \hspace{-0.5cm}
\includegraphics[width=0.52\textwidth]{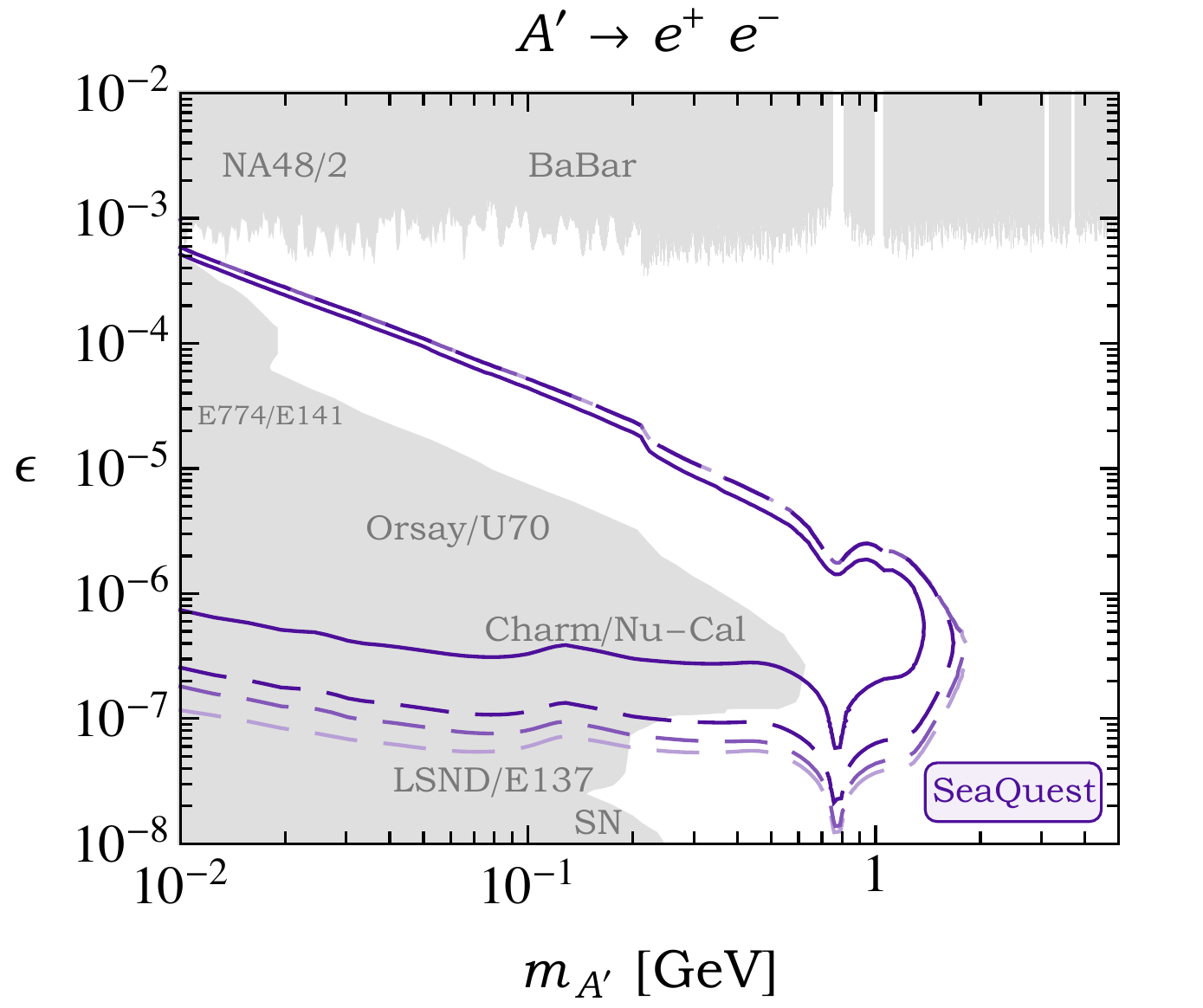} \hspace{-0.5cm}
\caption{\textbf{Left panel}: The projected Phase I SeaQuest sensitivity to the dark photon parameter space using the $5 \m - 6 \m$ fiducial decay region. The various contours correspond to 10 dielectron signal events for dark photons produced from meson ($\pi^0,\eta,\eta^\prime, \omega$) decays and proton Bremsstrahlung. The blue shaded region represents the theoretical uncertainty in computing the Bremsstrahlung rate (see text for details). \textbf{Right panel}: Seaquest sensitivity to displaced dark photons at Phase I (solid purple) and Phase II (dashed purple), corresponding to 10 signal events. For Phase I, we conservatively fix the fiducial decay region to $5 \m - 6 \m$. For Phase II, moving from darker to lighter contours corresponds to the fiducial decay regions of $5 \m - 6 \m$, $5 \m - 9 \m$, and $5 \m - 12 \m$, respectively. 
The gray region denotes parameter space that is already excluded by past experiments~\cite{Battaglieri:2017aum,Alexander:2016aln}.}
\label{fig:visAp}
\end{figure*}

Dark photons produced from proton Bremsstrahlung are generally more boosted than those originating from meson decays. This is seen in Fig.~\ref{fig:KinDarkPhoton}, which shows the energy and angular distributions of electrons originating from dark photon decays. Dark photons that are produced from Bremsstrahlung are much more energetic and strongly peaked in the forward direction. As a result, the electrons from their decays can more easily remain within the geometric acceptance of the spectrometer after passing through KMAG. This explains the overall hierarchy between the various efficiency lines in Fig.~\ref{fig:efficiencyDarkPhoton}.
However, for $\mAp \gtrsim \text{GeV}$, the acceptance for Bremsstrahlung is slightly reduced since these electrons often have sufficient $p_T \simeq E_e|\sin\theta_e|$ to escape the instrument geometry. In comparison, dark photons that are produced via meson decays are peaked at much smaller energies and at larger angles off of the beam-axis. The magnetic field of KMAG often sweeps these softer electrons out of the  spectrometer. In this case, the geometric acceptance is enhanced for energetic electrons originating from the decays of more massive dark photons, as seen in Fig.~\ref{fig:efficiencyDarkPhoton}. 

In the left panel of Fig.~\ref{fig:visAp}, we present the projected sensitivity of a Phase I displaced electron search to long-lived dark photons for the fiducial decay region of $5 \m - 6 \m$. We show separate contours corresponding to 10 signal events for the different production channels described in Sec.~\ref{sec:darkphoton}. Decays of heavier mesons, such as the $\eta$, contribute significantly to the dark photon reach for intermediate masses, $\mAp \sim 0.1 \GeV - 0.5 \GeV$, while the contribution from Drell-Yan is negligible. In the right panel of Fig.~\ref{fig:visAp}, we illustrate the projected SeaQuest reach both at Phase I (solid) and Phase II (dashed) after summing over the various production modes. For Phase I, we present results only for the minimal  $5 \m - 6 \m$ fiducial decay region. For Phase II, moving from the darker to lighter dashed contours corresponds to decay regions of $5 \m - 6 \m$, $5 \m -  9 \m$, and $5 \m - 12 \m$, respectively. 
We also compare the sensitivity of SeaQuest to existing constraints (gray) (see, e.g., Refs.~\cite{Battaglieri:2017aum,Alexander:2016aln} for a comprehensive review). 

\begin{figure}[t]
\hspace{-0.5cm}
\includegraphics[width=0.5\textwidth]{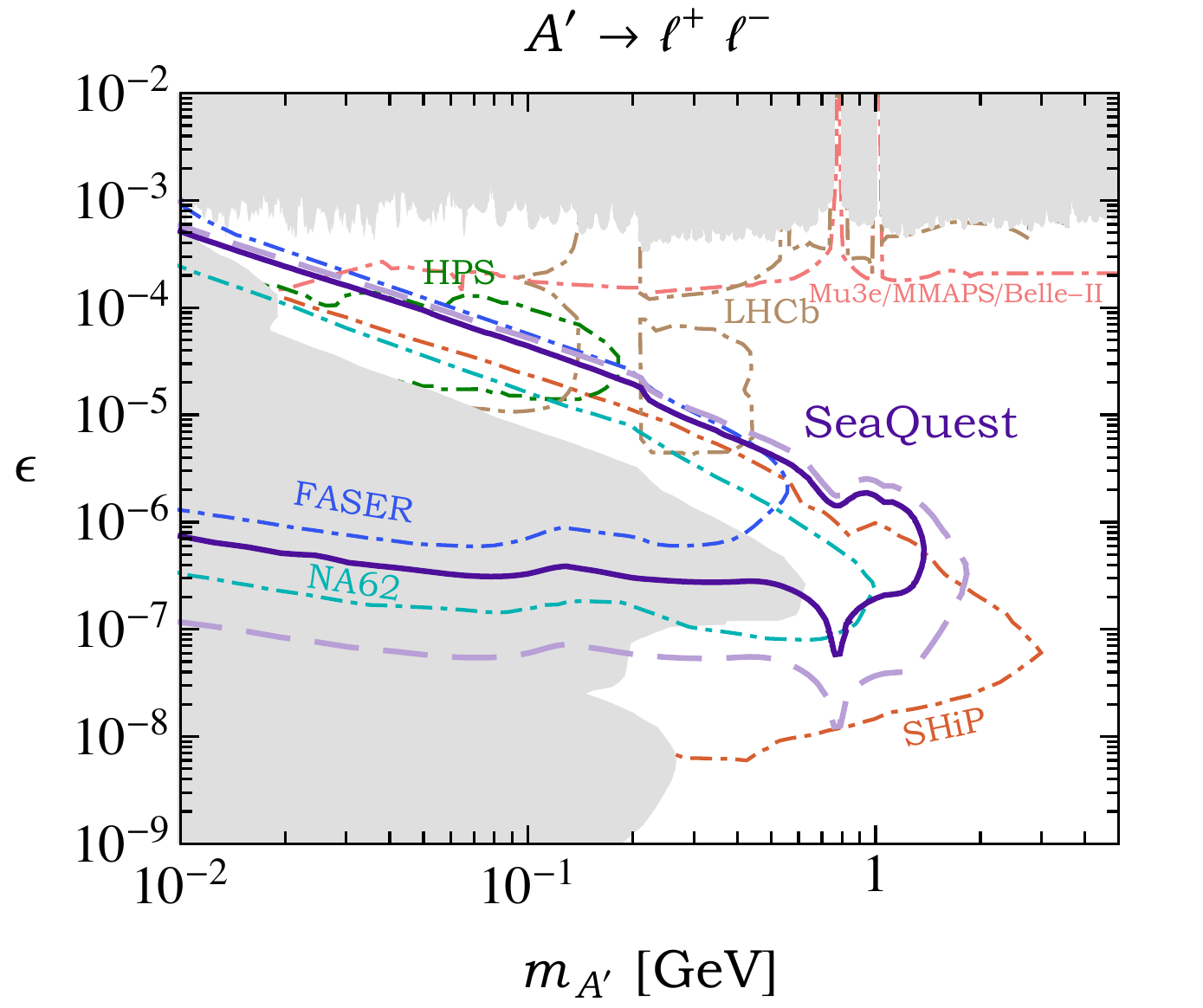} \hspace{-0.5cm}
\caption{As in the right panel of Fig.~\ref{fig:visAp}, the projected sensitivity of SeaQuest to displaced decays of dark photons at Phase I (solid purple) and Phase II (dashed purple) for the fiducial decay regions of $5 \m - 6 \m$ and $5 \m - 12 \m$, respectively, compared to existing constraints (solid gray)~\cite{Battaglieri:2017aum,Alexander:2016aln}. Also shown are the projected reach of the HPS (green) and Mu3e, MMAPS, and Belle-II experiments (pink)~\cite{Battaglieri:2014hga,Battaglieri:2017aum,Alexander:2016aln}, a beam dump run of NA62 (cyan)~\cite{na62cern}, and futuristic searches at LHCb (brown)~\cite{Ilten:2015hya,Ilten:2016tkc} and the proposed experiments FASER (blue)~\cite{Feng:2017uoz} and SHiP (red)~\cite{Alekhin:2015byh}.}
\label{fig:visApFuture}
\end{figure}

In Fig.~\ref{fig:visApFuture}, we highlight the ultimate reach of SeaQuest and compare it to other upcoming and proposed searches and experiments.  
The Heavy Photon Search (HPS) experiment~\cite{Battaglieri:2014hga,Alexander:2016aln} is sensitive to displaced leptonic decays of dark photons produced from electron-tungsten collisions. The green region in Fig.~\ref{fig:visApFuture} will be probed by HPS after accumulating  a $50 - 500$ nA current in the $1 - 6$ GeV energy range (this is expected by the end of 2018). The projected reach of LHCb is shown in brown, after accumulating $\sim 15$ fb$^{-1}$ of luminosity in Run 3~\cite{Ilten:2016tkc}: the region above the dimuon threshold could be explored by an inclusive dark photon search and the region below by a search for $D^{*0}\to D^0 ~ A^\prime (\to e^+ e^-)$. Also shown in cyan, blue, and red are the projected sensitivities of a beam dump run of NA62 after having accumulated $2\times 10^{18}$ POT~\cite{na62cern}, and the proposed FASER~\cite{Feng:2017uoz} and SHiP experiments at CERN~\cite{Alekhin:2015byh}, respectively. 

SeaQuest is capable of probing currently unexplored regions of parameter space. With Phase I luminosity and the ECAL upgrade, SeaQuest will explore dark photons up to $\sim 1.5 \GeV$ in mass, exceeding the mass reach of past proton fixed-target experiments, such as CHARM and Nu-Cal, and of the FASER proposal. Due to the relatively compact setup of the instrument, SeaQuest will also test larger values of kinetic mixing ($\epsilon \sim 10^{-5}$ for $\mAp \sim 100 \MeV$) that are challenging for longer-baseline experiments like SHiP or NA62. 
SeaQuest's ability to acquire data in the next few years with an already existing spectrometer highlights an obvious advantage compared to futuristic runs of much larger and costlier experiments in the coming decade. We also note that compared to previous projections (see, e.g., Ref.~\cite{Gardner:2015wea}), we find that SeaQuest will be sensitive to slightly larger values of $\epsilon$ and significantly larger values of $\mAp$, which is closer to the projections shown in, e.g., Ref.~\cite{mingmaryland}.

\section{Inelastic Dark Matter} 
\label{sec:iDM}
Additional $\Ap$ decay channels, beyond those of the minimal scenario discussed in Sec.~\ref{sec:SeaQuestAprime}, arise if the dark photon is not the lightest particle in the hidden sector. For instance, if there are new states that are directly charged under $U(1)_D$ and lighter than $\mAp / 2$, then dark photon decays to the hidden sector naturally dominate over those to SM species provided that $\alpha_D \gg \alpha_\text{em} \epsilon^2$, where $\alpha_D \equiv e_D^2 / 4 \pi$ and $e_D$ is the $U(1)_D$ gauge coupling.
In this case, a plethora of different signatures is accessible at SeaQuest and other fixed-target experiments. For example, if the lighter species is stable on collider timescales, invisible $\Ap$ decays can be searched for at low-energy missing energy and momentum experiments~\cite{Battaglieri:2017aum,Alexander:2016aln}. Less minimal models involving additional particles in the dark sector often predict longer decay chains, and $\Ap$ production leads to several visible and invisible particles in the final state~\cite{Morrissey:2014yma}. Such a study of SeaQuest within the context of strongly interacting DM has recently appeared in Ref.~\cite{Berlin:2018tvf}. In this section, we investigate similar types of experimental signatures in models of inelastic DM (see Refs.~\cite{Izaguirre:2015zva,Izaguirre:2017bqb} for recent studies of similar phenomenology).

\subsection{Model}
\label{eq:iDMmodel}

Models of inelastic DM (iDM) were first proposed as a viable explanation to the longstanding DAMA anomaly~\cite{TuckerSmith:2001hy} and have continued relevance in a broader parameter space independent of this anomaly.  The essential physics of iDM is that in some models (such as DM interacting through a massive vector mediator) it is generic for DM to (1) have two nearly (but not exactly) degenerate mass states, and (2) interact primarily through mass-off-diagonal couplings.  In this section, we discuss not the DAMA-motivated parameter space of iDM but the generic physics described above, which is an attractive framework for GeV-scale thermal DM. It is well-known that thermal DM lighter than $\sim 10 \GeV$ must have suppressed annihilations at late times in order to alleviate strong bounds from measurements of the cosmic microwave background (CMB)~\cite{Ade:2015xua}. This is accomplished within models of iDM since freeze-out dominantly occurs through coannihilations of the DM with its slightly heavier counterpart, whose population is sufficiently depleted at the time of recombination. As a result,  late-time annihilations are suppressed below detectable levels.  Furthermore, since iDM naturally has suppressed scattering rates in underground direct detection experiments,\footnote{See, however, Ref.~\cite{Bramante:2016rdh} for novel venues to test iDM at direct detection experiments.} dedicated searches at low-energy accelerators constitute a prime avenue towards the detection of light DM in this class of models.

We focus on the particular implementation of iDM involving a single Dirac pair of two-component Weyl spinors, $\eta$ and $\xi$, oppositely charged under the broken $U(1)_D$ symmetry. Similar to Sec.~\ref{sec:darkphoton}, we assume that the dark photon associated with $U(1)_D$, $\Ap$, kinetically mixes with SM hypercharge (see Eq.~(\ref{eq:KinMix})). In addition to the Dirac mass, $m_D$, allowed by all symmetries of the model, it is also natural to include $U(1)_D$-breaking Majorana mass terms, $\delta_{\eta, \xi}$, for each Weyl component.  These are naturally generated by the same $U(1)_D$-breaking spurion that is responsible for generating the $\Ap$ mass (such as a dark Higgs). We therefore take as our simplified Lagrangian,
\beq
- \mathcal L \supset m_D \, \eta \, \xi + \frac{1}{2} \, \delta_\eta \, \eta^2 + \frac{1}{2} \, \delta_\xi \, \xi^2 + {\text{h.c.}}
 \eeq
Since $\delta_{\eta, \xi}$ explicitly breaks $U(1)_D$, it is technically natural to take $ \delta_{\eta, \xi} \ll m_D$. Hereafter, we will adopt this limit, in which case the spectrum consists of a pseudo-Dirac pair of nearly degenerate Majorana fermions that couple off-diagonally (inelastically) to the $\Ap$. 

In the physical mass basis, the eigenvectors, denoted by $\chi_1$ and $\chi_2$, have a mass given by
\beq
m_{1,2} \simeq m_D \mp \frac{1}{2} (\delta_\eta+\delta_\xi)
~,
\eeq
where
\begin{align}
\chi_1 &\simeq i (\eta - \xi) / \sqrt{2} 
\nl
\chi_2 &\simeq (\eta + \xi)/\sqrt{2}
~.
\end{align}
The lightest state, $\chi_1$, is cosmologically stable and can constitute a DM candidate. The hierarchy $\delta_{\eta, \xi} \ll m_D$ translates into a small fractional mass splitting, 
\be
\Delta\equiv \frac{m_2-m_1}{m_1} \simeq \frac{\delta_\eta + \delta_\xi}{m_D} \ll 1
~.
\ee
Hereafter, we will focus on mass splittings of size $\Delta \sim 0.1$, where $\Delta$ is not large enough to dramatically affect the cosmological DM relic abundance, but decays of $\chi_2$ into $\chi_1$ lead to detectable signals, as we will describe below.
In four-component notation, the two Majorana fermions, $\chi_{1,2}$, interact with $A^\prime$ dominantly through the inelastic interaction
\beq
\mathcal{L} \supset ie_D ~ A^\prime_\mu ~ \bar\chi_1\gamma^\mu\chi_2
~.
\eeq
In general, there is also an elastic coupling of the form
\beq
\mathcal{L} \supset \frac{
e_D ~ ( \delta_\eta - \delta_\xi)}{4 \, m_D}~ A^\prime_\mu ~(\bar\chi_1\gamma^\mu \gamma^5 \chi_1-\bar\chi_2\gamma^\mu \gamma^5 \chi_2)
~.
\eeq
Note that this term vanishes for $\delta_\eta = \delta_\xi$, since these interactions violate the enhanced charge-conjugation symmetry, $\Ap \leftrightarrow - \Ap$ and $\eta \leftrightarrow \xi$ (equivalently, $\chi_{1,2} \leftrightarrow \mp \chi_{1,2}$).  
In the more general case where $\delta_\eta \neq\delta_\xi$, the elastic coupling is non-zero but naturally suppressed (compared to the inelastic piece) by the small ratio of Majorana and Dirac masses. Throughout this work, we assume that contributions from elastic interactions are negligible.

In this framework, decays of the $\Ap$ to SM fermions are suppressed 
by $\epsilon^2 \ll 1$ (see  Eqs.~(\ref{eq:ApDecaytoLeptons}) and (\ref{eq:ApDecaytoHadrons})). On the other hand, $\chi_{1,2}$ couplings to the $\Ap$ are proportional to $e_D$ with no $\epsilon$-suppression. Hence, if $\alpha_D \gg \alpha_\text{em} \epsilon^2$ and $\mAp > m_1 + m_2$, then the dark photon decays almost exclusively to a $\chi_1\, \chi_2$ pair. The corresponding partial width is given by 
\be
\Gamma (\Ap \to \chi_1 \chi_2) \simeq \frac{\alpha_D \mAp}{3} \sqrt{ 1 - \, \frac{4m_1^2}{\mAp^2}} 
 \left( 1 +  \, \frac{2m_1^2}{\mAp^2} \right)
\ee
where we have taken $\Delta \ll 1$ ($m_1 \simeq m_2$). We will henceforth assume the theoretically motivated hierarchies $\alpha_D \gg \alpha_\text{em} \epsilon^2$, $\Delta\ll 1$, as well as $m_1 < \mAp$, which is cosmologically motivated for light DM as explained below.

\begin{figure}[t]
\begin{center}
\includegraphics[width=0.5\textwidth]{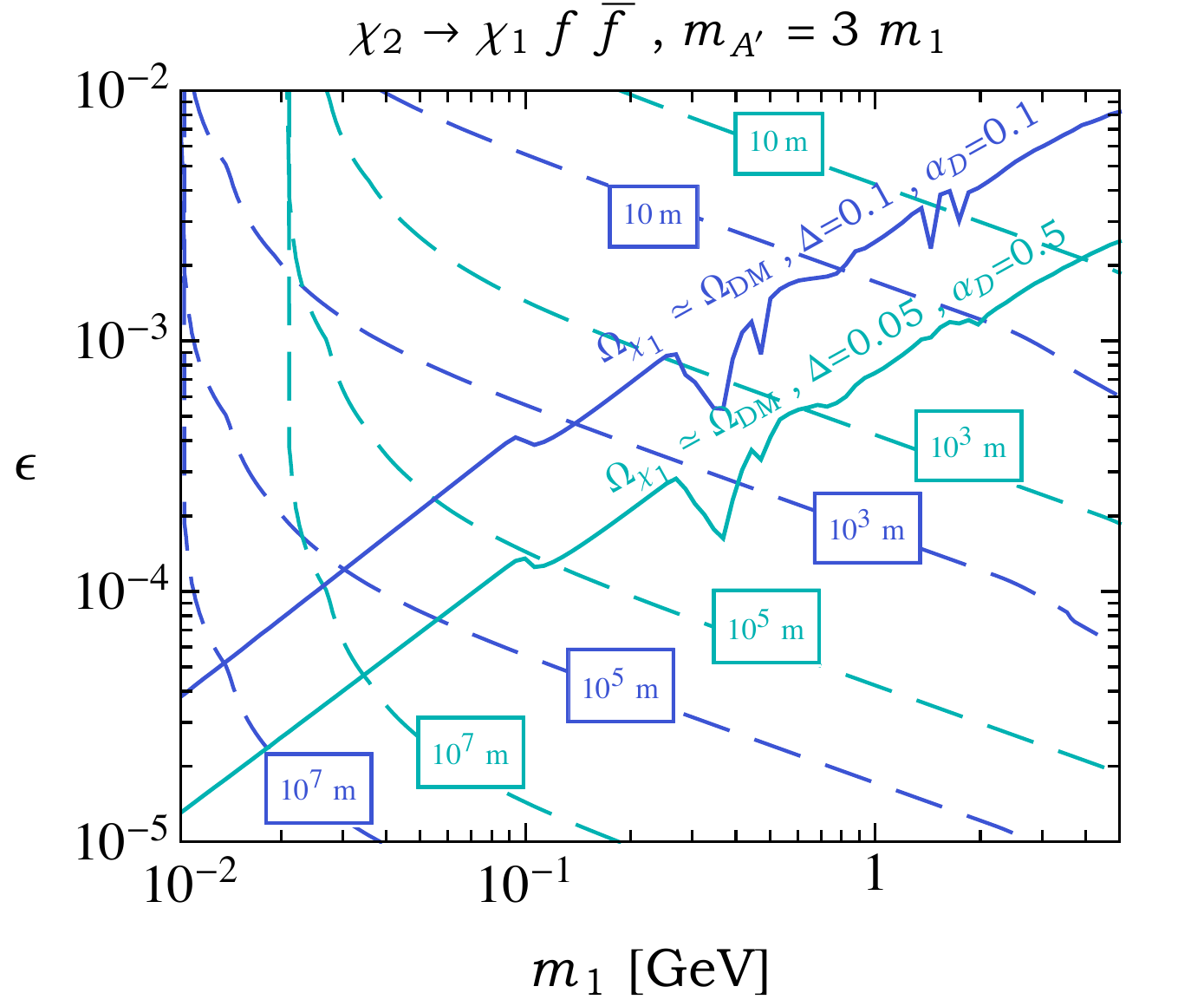}
\caption{The proper lifetime of the excited state, $\chi_2$, for $\Delta=0.1$, $\alpha_D=0.1$ (dashed blue) and $\Delta=0.05$, $\alpha_D=0.5$ (dashed cyan) fixing $\mAp=3 \, m_1$. 
Also shown for these two benchmarks (solid blue and solid cyan, respectively) are contours corresponding to regions of parameter space where the abundance of $\chi_1$ agrees with the measured DM energy density. Below the corresponding lines, $\chi_1$ is overabundant.}\label{fig:Chi2_Decay}
\end{center}
\end{figure}

In the standard WIMP paradigm, DM is assumed to be in equilibrium with the SM bath at large temperatures, $T \gg m_{1,2}, \, \mAp$. The dominant equilibrating processes are often decays and inverse-decays of the dark photon, $\Ap \leftrightarrow f \bar{f}$, where $f$ is an electrically charged SM fermion. 
Thermal equilibrium between the SM and hidden sector baths is guaranteed if the rate for such processes, $\Gamma \sim \alpha_\text{em} \, \epsilon^2 \left( \mAp^2 / T \right)$, exceeds the Hubble parameter, $H \sim T^2 / m_\text{pl}$. Hence, demanding that kinetic equilibration occurs before $T \lesssim m_1$ sets a lower bound on $\epsilon$ that approximately scales as
\be
\label{eq:KE}
\epsilon \gtrsim \order{10^{-8}} \times \left(\frac{m_1}{\text{GeV}}\right)^{1/2}
~,
\ee
where we have taken $m_1 \sim \mAp$. For values of $\epsilon$ that exceed this lower bound, a large thermal population of $\chi_{1,2}$ is necessarily generated in the early universe, which must be sufficiently depleted at late times. 

For $m_1 > \mAp$,  DM freeze-out dominantly proceeds through direct annihilations into pairs of dark photons, $\chi_1 \chi_1 \to \Ap \Ap$, followed by $\Ap \to f \bar{f}$~\cite{Pospelov:2007mp}. For $m_1 \lesssim 10 \GeV$, such processes are in conflict with measurements of the CMB since the corresponding annihilation rate is unsuppressed at low velocities~\cite{Ade:2015xua}. If, on the other hand, $m_1 < \mAp$, DM freeze-out is dictated through coannihilations involving an intermediate $\Ap$, i.e., $\chi_1 \chi_2 \to A^{\prime *} \to f \bar{f}$. This process is exponentially suppressed by the relative mass splitting between $\chi_2$ and $\chi_1$, i.e., $\sim \text{exp}\left[ - \Delta \, m_1 / T \right]$~\cite{Griest:1990kh}. A simple parametric estimate shows that the abundance of $\chi_1$ is in agreement with the observed DM energy density for
\be
\label{eq:thermalm1}
m_1 \sim \frac{\epsilon \, \left( \alpha_D \, \alpha_\text{em} \, T_\text{eq} \, m_\text{pl} \right)^{1/2}}{\left( \mAp / m_1 \right)^2} ~ e^{ -x_f \Delta / 2 }
~,
\ee
where $T_\text{eq} \simeq 0.8 \eV$ is the temperature at matter-radiation equality, and $x_f \equiv m_1 / T \sim \order{10}$ at freeze-out. Hence, for sufficiently small mass splittings ($\Delta \lesssim 0.1$) freeze-out proceeds as in the standard WIMP-like manner, but is exponentially suppressed at the time of recombination, significantly relaxing the strong CMB constraints. Larger values of the fractional mass splitting, $\Delta$, are also viable but require larger values of $\epsilon$ or $\alpha_D$ for fixed $\chi_1$ and $\Ap$ masses~\cite{DAgnolo:2018wcn}. 

In our analysis, we numerically calculate the relic abundance of the $\chi_1$ population, following the procedure outlined in Refs.~\cite{Griest:1990kh,Edsjo:1997bg} and including hadronic final states as discussed in Ref.~\cite{Izaguirre:2015zva}. Along the solid blue (cyan) contour of Fig. \ref{fig:Chi2_Decay}, the relic abundance of $\chi_1$ agrees with the measured DM energy density for $\mAp=3 \, m_1$, $\Delta=0.1~(0.05)$, and $\alpha_D=0.1 ~ (0.5)$. For specific values of $m_1$, annihilations at freeze-out are significantly enhanced through resonant mixing of the intermediate $\Ap$ with SM vector mesons, analogous to $\gamma - \rho^0$ mixing in the SM. We incorporate the effect of these spin-1 resonances through the data-driven parameter $R(\sqrt{s})$ (see Eq.~(\ref{eq:ApDecaytoHadrons})). In addition to the $\rho^0$, $\omega$, $\phi$, and $\rho^\prime$ mesons, we have manually included the contributions from the narrow $J/\psi$, $\psi(2 S)$, and $\Upsilon$ resonances. We find that thermal averaging of the DM annihilation rate significantly suppresses these latter contributions.\footnote{Compared to the relic abundance calculation in Ref.~\cite{Izaguirre:2015zva}, we have implemented the physical widths of the spin-1 hadronic resonances in the numerical form of $R(\sqrt{s})$. In the case of the heavier narrow resonances, we find that this significantly reduces the effect of $\Ap$-meson mixing on the calculation of the $\chi_1$ relic abundance.}

\begin{figure*}[t]
\hspace{-0.5cm}
\includegraphics[width=0.5\textwidth]{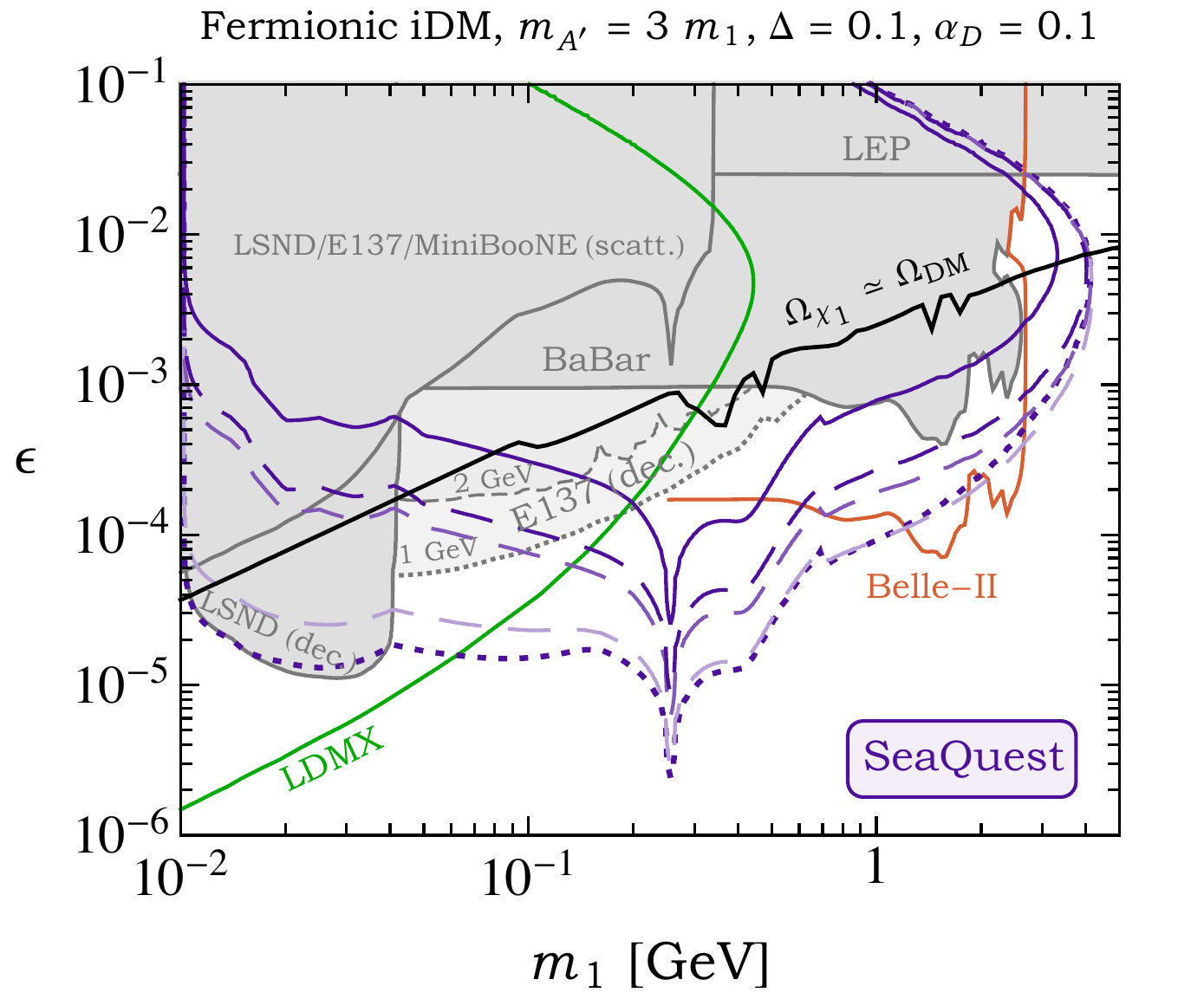} \hspace{-0.5cm}
\includegraphics[width=0.5\textwidth]{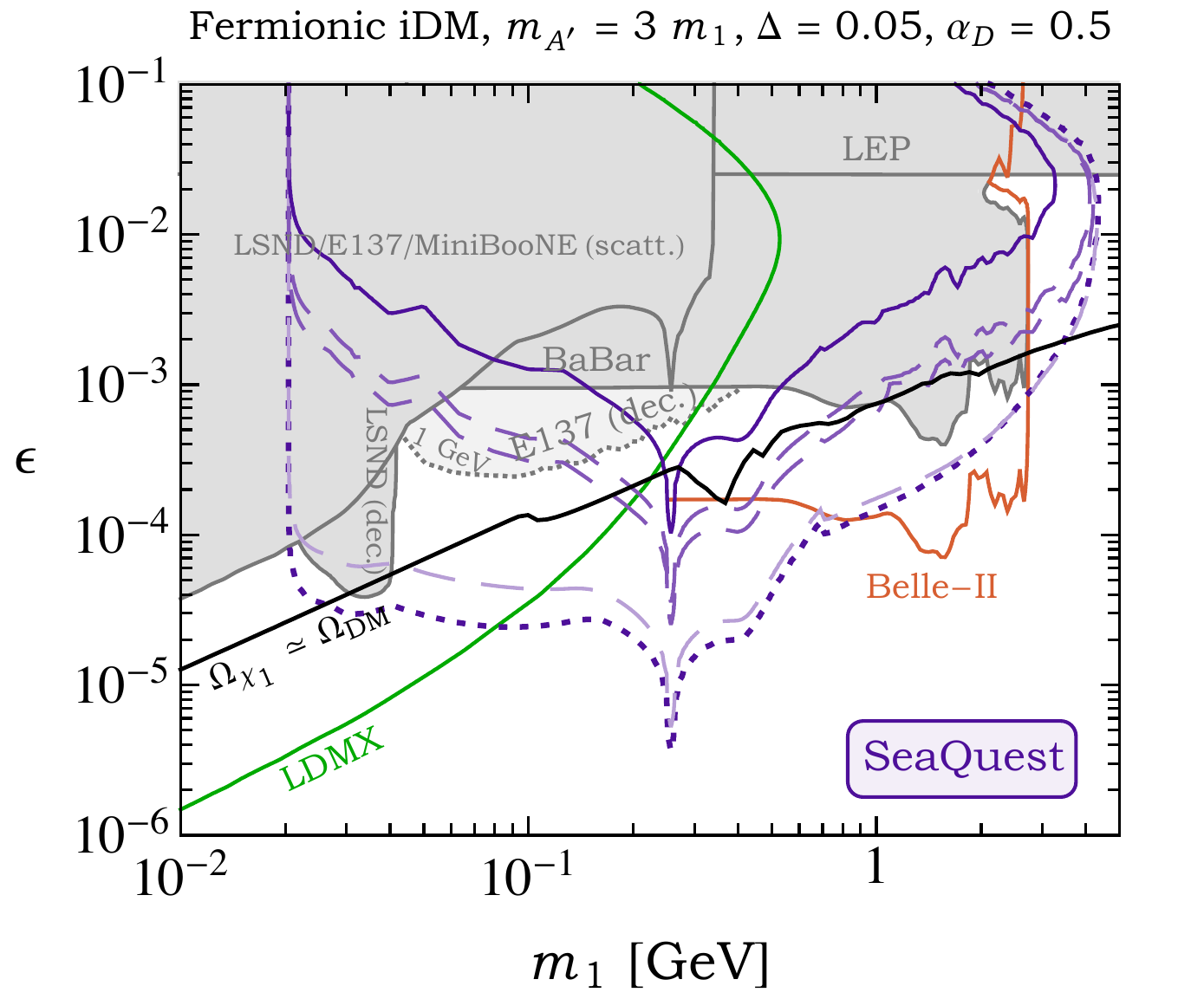} \hspace{-0.5cm}\\
\caption{Existing constraints (shaded gray) and projected sensitivities (color) to models of fermionic inelastic dark matter in the $m_1 - \epsilon$ plane. In each panel, we have fixed $\mAp / m_1 = 3$, while in the left (right) panel we take $\Delta = 0.1$ (0.05) and $\alpha_D = 0.1$ (0.5). Along the black contour, the abundance of $\chi_1$ matches the observed dark matter energy density. The shaded regions are excluded by LEP~\cite{Hook:2010tw,Curtin:2014cca}, BaBar~\cite{Aubert:2008as, Lees:2017lec}, dark matter scattering at LSND~\cite{deNiverville:2011it,Auerbach:2001wg}, E137~\cite{Batell:2014mga,Bjorken:1988as}, and MiniBooNE~\cite{Aguilar-Arevalo:2017mqx}, and visible signals of decays at E137~\cite{Bjorken:1988as} and LSND~\cite{Athanassopoulos:1996ds}. For visible decay signals at E137, the dotted (dashed) gray contours correspond to an energy deposition threshold of 1 GeV (2 GeV). The colored lines correspond to the projected reach of Belle-II (orange)~\cite{Essig:2013vha,Battaglieri:2017aum,Alexander:2016aln}, LDMX (green)~\cite{Izaguirre:2014bca}, and SeaQuest (purple), as described in the text. The projected reach of SeaQuest is shown as in Fig.~\ref{fig:visAp}. For Phase I (solid purple), we conservatively fix the fiducial decay region to $5 \m - 6 \m$. For Phase II (dashed purple), moving from darker to lighter contours corresponds to the fiducial decay regions: $5 \m - 6 \m$, $5 \m - 9 \m$, and $5 \m - 12 \m$, respectively. We also show the Phase II reach for the $5 \m - 6 \m$ decay region, assuming that the electrons do not have to travel through the magnetic field of KMAG (dotted purple).}
\label{fig:iDMSummary}
\end{figure*}

As shown in Fig.~\ref{fig:Chi2_Decay}, for sizable values of $\alpha_D$ and DM masses not much heavier than a few GeV, the measured relic abundance is reproduced for $\epsilon^2 \ll \alpha_D$, favoring an $\order{1}$ branching ratio for $\Ap \to \chi_1 \, \chi_2$. 
Such decays constitute the dominant production mechanism for DM, $\chi_1$, and its excited state, $\chi_2$, if light dark photons are produced at accelerators, as in Sec.~\ref{sec:darkphoton}.  
Once produced, $\chi_2$ subsequently decays back to SM fermions through an off-shell dark photon, i.e., $\chi_2 \to \chi_1 \, A^{\prime *}\to \chi_1 \, f \bar f$. 
In the limit that $\mAp \gg m_1$, $\Delta \ll 1$, and $m_\ell \simeq 0$, 
the partial width into $\chi_1$ and a single pair of SM leptons is given by
\beq\label{eq:decaychi2}
\Gamma(\chi_2\to\chi_1 \, \ell^+\ell^-)\simeq \frac{4\, \epsilon^2\, \alpha_\text{em} \, \alpha_D \, \Delta^5 m_1^5}{15\pi \, \mAp^4}
~.
\eeq
In most of the parameter space of interest for SeaQuest with $\Delta \lesssim 0.1$, decays to electrons dominate. However, we have included decays to muons and hadrons when kinematically allowed (with the latter contributing at most $\order{10} \%$).
The decay rate in Eq.~(\ref{eq:decaychi2}) implies that $\chi_2$ is long-lived on collider length-scales for GeV-scale masses and $\order{0.01} -\order{0.1}$ mass splittings. This is demonstrated in Fig. \ref{fig:Chi2_Decay}, where we additionally show the proper lifetime of $\chi_2$ as dashed contours for $\mAp=3 \, m_1$ and for various values of $\Delta$ and $\alpha_D$, as a function of $m_1$ and $\epsilon$. The blue (cyan) contours correspond to $\Delta=0.1$ ($\Delta=0.05$) and $\alpha_D=0.1$ ($\alpha_D=0.5$).
In the cosmologically motivated regions of parameter space, the $\chi_2$ lifetime is typically much greater than $\sim 1 \m$. 
These considerations open up the possibility of experimental searches for visible displaced decays of $\chi_2$.

\begin{figure*}[t]
\hspace{-0.5cm}
\includegraphics[width=0.5\textwidth]{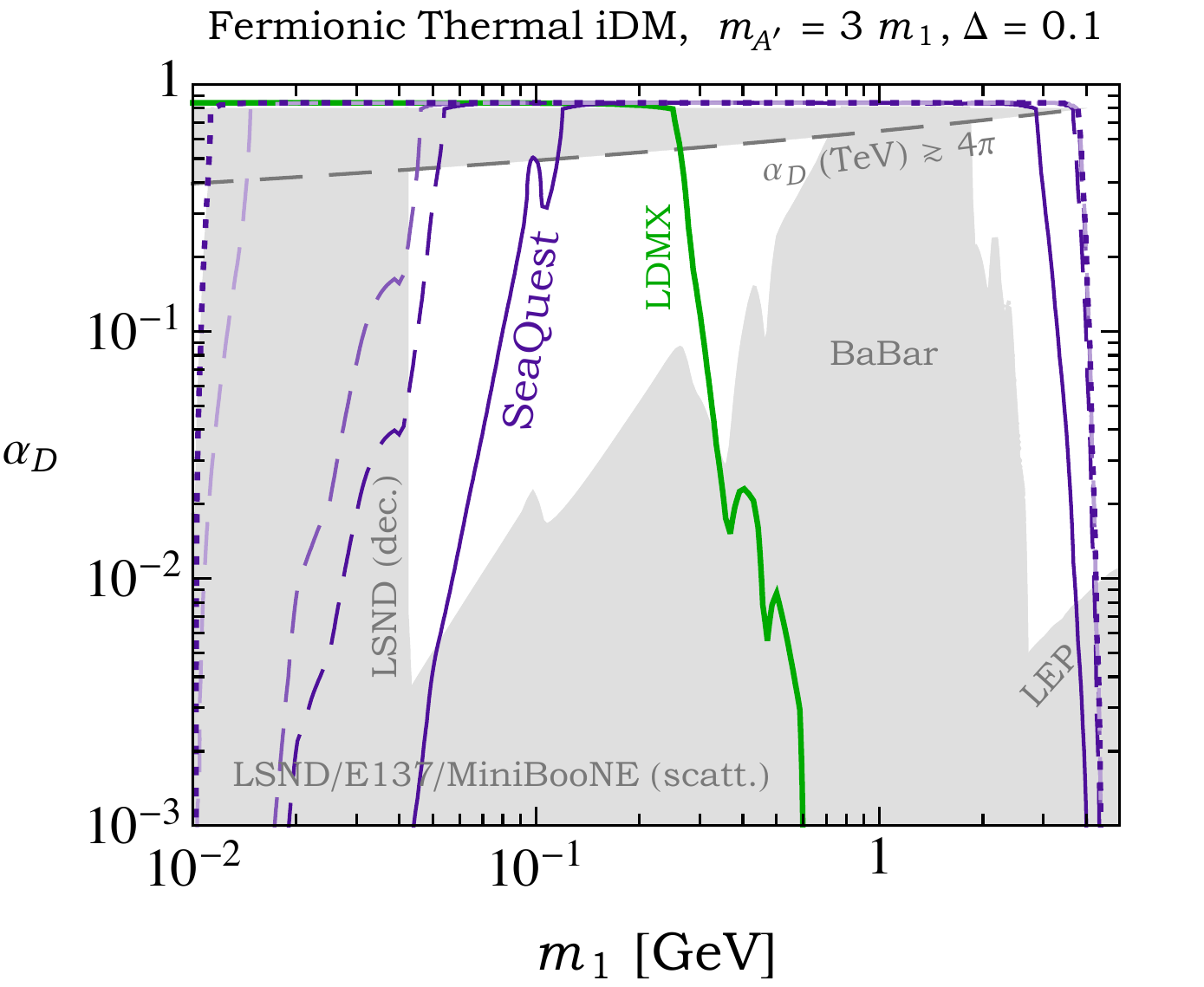} \hspace{-0.5cm}
\includegraphics[width=0.5\textwidth]{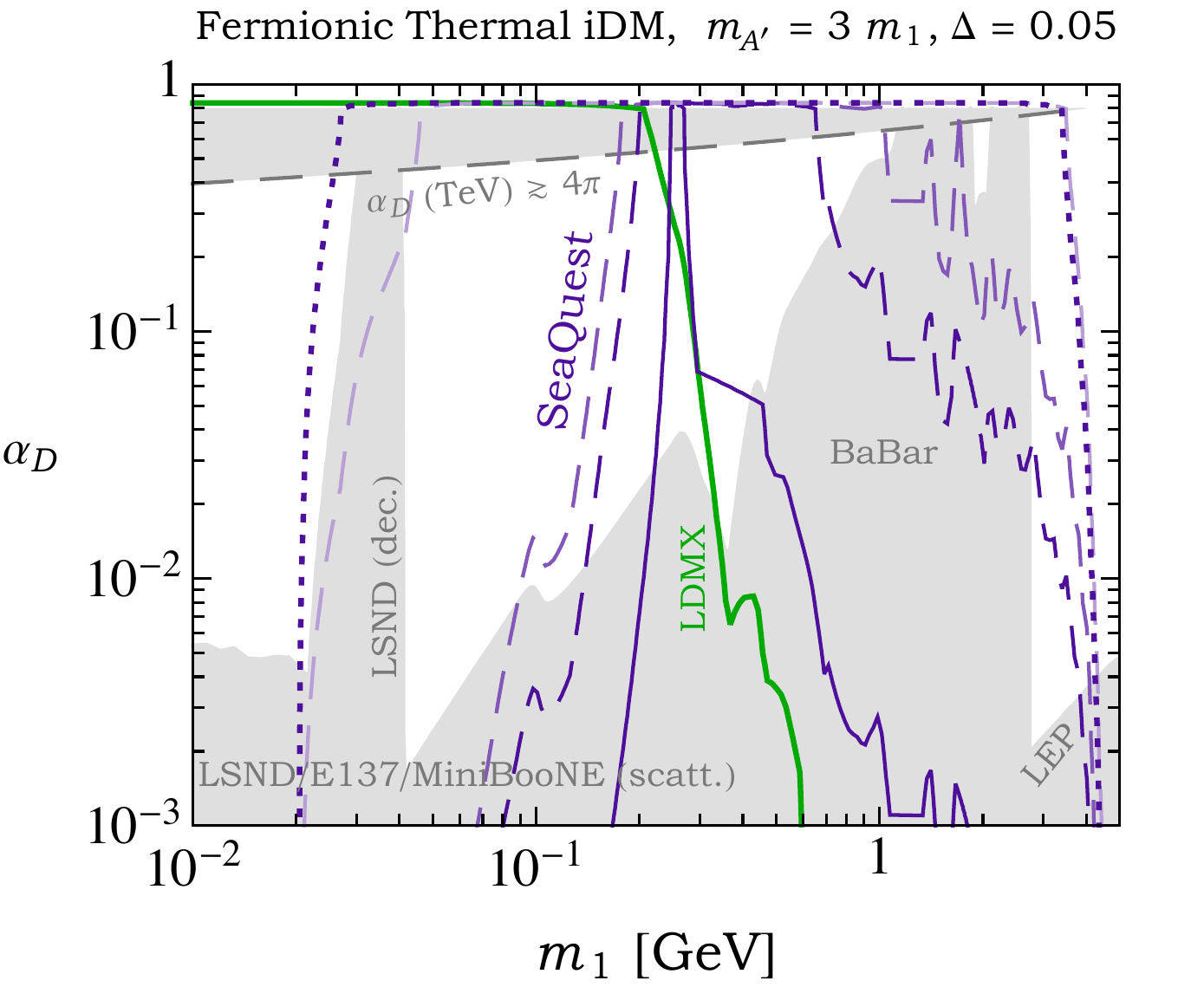} \hspace{-0.5cm}
\caption{As in Fig.~\ref{fig:iDMSummary}, existing constraints (shaded gray) and projected sensitivities (color) to models of thermal inelastic dark matter in the $m_1 - \alpha_D$ plane. For each point in parameter space, the kinetic mixing parameter, $\epsilon$, is fixed such that $\chi_1$ freezes out with an abundance that is in agreement with the observed dark matter energy density. Above the gray dashed line, $\alpha_D$ becomes non-perturbative at $\sim 1 \TeV$. An artificial cutoff is added to the panels at $\alpha_D \simeq 0.8$ as a visual aid.}
\label{fig:iDMthermal}
\end{figure*}
%

\subsection{Review of Existing Constraints}
\label{sec:iDMconstraints}

There is currently an extensive program involving beam dump, fixed-target, and collider experiments in the search for new physics below the GeV-scale~\cite{Alexander:2016aln,Battaglieri:2017aum}. In fact, existing searches are already sensitive to the DM models discussed above. In this section, we review the multitude of constraints on various signals inherent to models of light iDM. Figs.~\ref{fig:iDMSummary} and \ref{fig:iDMthermal} illustrate the viable iDM parameter space, fixing $\mAp / m_1 = 3$. In Fig.~\ref{fig:iDMSummary}, we show existing exclusions (shaded gray) and projected sensitivities (color) of experimental searches in the $m_1 - \epsilon$ plane for two benchmark values of $\Delta$ and $\alpha_D$. Along the black contour, the abundance of $\chi_1$ matches the observed DM energy density. In Fig.~\ref{fig:iDMthermal}, we examine the restricted \emph{thermal} iDM parameter space in the $m_1 - \alpha_D$ plane. For each point, we fix $\epsilon$ to the value required for $\chi_1$ to freeze out with an adequate cosmological abundance. $\alpha_D$ is assumed to be the value defined at the scale $\mu \sim m_1$. We demand that $\alpha_D$ is perturbative after RG evolving up to the weak scale. Above the gray dashed line, $\alpha_D$ is non-perturbative at the scale $\mu \sim 1 \TeV$ ($\alpha_D(\rm{TeV})\gtrsim 4\pi$).

\textbf{High-energy colliders:}
Experiments that rely solely on the dark photon's coupling to the SM, independent of its decay modes and any other interactions in the hidden sector, constitute highly model-independent probes of these theories. Kinetic mixing between the dark photon and the SM hypercharge gauge boson generically leads to a shift in several electroweak precision observables as measured by LEP.  
A recent fit to a set of relevant observables 
leads to the constraint $\epsilon\lesssim 3 \times 10^{-2}$~\cite{Hook:2010tw,Curtin:2014cca}. This is approximately independent of the mass of the dark photon, as long as  $\mAp\ll m_Z$.

\textbf{B-factories:}
At accelerators, searches for dark photon decays are sensitive to additional couplings in the hidden sector. For instance, the BaBar collaboration has performed a direct search for invisibly decaying dark photons, i.e., $e^+ e^- \to \gamma + \Ap(\to \text{invisible})$~\cite{Aubert:2008as, Lees:2017lec}. 
The reach does not strongly depend on the exact value of $\alpha_D$ 
provided that $\alpha_D \gg \alpha_\text{em} \epsilon^2$, since in this case the dark photon decays almost exclusively to $\chi_1\chi_2$ pairs. 
Such events are signal-like if the decay $\chi_2 \to \chi_1 \ell^+ \ell^-$ occurs outside of the detector or the soft lepton pair falls below the detector thresholds. In recasting the limits of Ref.~\cite{Aubert:2008as}, we demand that $\chi_2$ travels a radial length of $1.5 \m$ before decaying or that $\chi_2$ decays inside the detector to leptons that are too soft to be detected, $p_T^{\ell^\pm} \lesssim 60 \MeV$~\cite{Aubert:2001tu}. In the more recent BaBar study of Ref.~\cite{Lees:2017lec}, a larger dataset was acquired. In recasting this analysis, incorporating a similar lepton veto is not as straightforward, as signal events are selected by a multivariate Boosted Decision Tree discriminant. Therefore, in rescaling these limits, we simply demand that $\chi_2$ decays outside of the detector. For each point in parameter space, we take the stronger constraint from either of these two recasted searches~\cite{Aubert:2001tu,Lees:2017lec}. 
Belle-II is expected to acquire $\sim 50 \text{ ab}^{-1}$ of data by the year $2023$ and could perform a similar mono-photon search, enhancing the sensitivity to $\epsilon$ by roughly a factor of $5-6$~\cite{Essig:2013vha,Battaglieri:2017aum,Alexander:2016aln}. The orange contour in Fig.~\ref{fig:iDMSummary} shows the estimated bound.  We note that this search is expected to become limited for $\mAp \lesssim \text{GeV}$, as it relies on the careful rejection and measurement of SM background. Hence, we do not explicitly show the Belle-II reach for these small masses. 

\textbf{Beam-dumps (scattering):}
As discussed in Sec.~\ref{sec:darkphoton}, dark photons can be directly produced at proton beam dump experiments. Similar processes can lead to a sizable flux of dark photons at electron beam experiments as well. If the $\Ap$ decays to long-lived DM states, a collimated DM beam can be produced at existing low-energy beam dumps. This energetic beam of DM particles can then be observed if it relativistically scatters (through $\Ap$ exchange) with electrons or nucleons in a detector placed downstream of the target. Strong constraints on light DM have been obtained from measurements performed at LSND~\cite{deNiverville:2011it,Auerbach:2001wg}, E137~\cite{Batell:2014mga,Bjorken:1988as}, and MiniBooNE~\cite{Aguilar-Arevalo:2017mqx}. 
In recasting these searches, we have simply rescaled the published bounds by the appropriate choice of $\alpha_D$. 

\textbf{Beam-dumps (decay):}
Beam dumps are also sensitive to the visible decays of the excited state. If $\chi_2$ is sufficiently long-lived (see Fig.~\ref{fig:Chi2_Decay}), dark photon production followed by $\Ap \to \chi_1 \chi_2 \to \chi_1 \chi_1 \ell^+ \ell^-$ leads to displaced leptons that can deposit observable energy into detectors  
at existing experiments. 

The 800 MeV proton beam at the Liquid Scintillator Neutrino Detector (LSND) at Los Alamos produced $\sim \order{10^{22}}$ neutral pions after running from 1993-1998~\cite{Athanassopoulos:1996ds}. From this large collection of pions, a huge number of dark photons may have been produced via $\pi^0 \to \gamma \Ap$ for $\mAp \lesssim 100 \MeV$. An off-axis scintillator detector was placed $\sim 30 \m$ downstream of the water-copper target, with sensitivity to energy depositions below $\sim 100 \MeV$, which could arise from the visible products of $\chi_2$ decays.  In estimating the rate of these events and extracting a constraint, we closely follow the analysis in Refs.~\cite{Essig:2010gu} and \cite{Izaguirre:2017bqb}, utilizing the GEANT pion simulation from Ref.~\cite{Kahn:2014sra} and manually decaying these pions to on-shell $\Ap$ final states as described in Sec.~\ref{sec:darkphoton}. 
In recasting these limits, we find good agreement with the results of Ref.~\cite{Izaguirre:2017bqb}. 

The E137 experiment at SLAC~\cite{Bjorken:1988as} was designed to look for displaced visible decays of light axions, produced from a 20 GeV electron beam impinging on a water-aluminum target. The experiment acquired an impressive amount of data, corresponding to roughly 30 C of current, equivalent to $\sim 10^{20}$ electrons on target (EOT) and an effective integrated luminosity of $\sim 100$ ab$^{-1}$. A $\sim 1 \m^3$ ECAL was placed $\sim 400 \m$ downstream of the aluminum target with $179 \m$ and $204 \m$ composed of natural shielding (in the form of a dirt hill) and an open-air decay region, respectively. Timing and geometric cuts effectively suppressed contributions from cosmic rays and sky shine, resulting in a background-free search.  

At this experiment, dark photons may have been produced through electron Bremsstrahlung. If the $\chi_2$ from the dark photon decay is sufficiently long-lived, it can traverse the dirt hill before decaying to electrons in the open decay region. We have simulated this process through a modified version of {\tt MadGraph5}~\cite{Alwall:2011uj} after implementing the iDM model in {\tt FeynRules}~\cite{Alloul:2013bka}. In recasting the E137 sensitivity,  we demand that an energetic electron from $\Ap \to \chi_1 \chi_2 \to \chi_1 \chi_1 e^+ e^-$ passes directly through the ECAL and points back to the target within an angular resolution of $\sim \order{10} \text{ mrad}$. We additionally model energy loss as the electrons traverse the open air, corresponding to a radiation length of $\sim 304 \text{ meters}$~\cite{Patrignani:2016xqp,Tsai:1966js}. 

 In the original data analysis of Ref.~\cite{Bjorken:1988as}, different thresholds were imposed on the energy deposited in the ECAL, ranging from $1 \GeV - 3 \GeV$. As distinct analyses in Ref.~\cite{Bjorken:1988as} utilized different thresholds, it is unclear which choice should be adopted in recasting these limits. The precise value of this threshold is not important for minimal models of visibly decaying axions or dark photons since the energy of the electrons from such decays is comparable to $E_\text{beam} = 20 \GeV$. However, in models of iDM, the longer decay chain and the small $\chi_2 - \chi_1$ mass splitting result in comparatively softer electrons from $\chi_2 \to \chi_1 e^+ e^-$ decays. The corresponding bound for iDM is therefore particularly sensitive to the uncertainty in the energy threshold. We therefore recast these limits for various energy threshold choices. Compared to Ref.~\cite{Izaguirre:2017bqb}, which only considered decays of $\chi_2$ inside the ECAL, we include decays throughout the entire open region leading up to the detector. 
The relevant regions of parameter space, shown as the outlined gray regions in Fig.~\ref{fig:iDMSummary}, correspond to model parameters where at least 3 signal events are expected during the run of this experiment. 
The dotted (dashed) gray contours correspond to recasted E137 bounds assuming an energy deposition threshold of 1 GeV (2 GeV); for a threshold of 3 GeV, E137 excludes none of the parameter space shown. We therefore conclude that $\chi_2$ decays in E137 do not robustly exclude any relevant parameter space for the models considered in Figs.~\ref{fig:iDMSummary}, and we do not show E137 decay exclusions in Fig.~\ref{fig:iDMthermal}.

\begin{figure*}[t] 
\vspace{0.cm}
\includegraphics[width=0.49\textwidth]{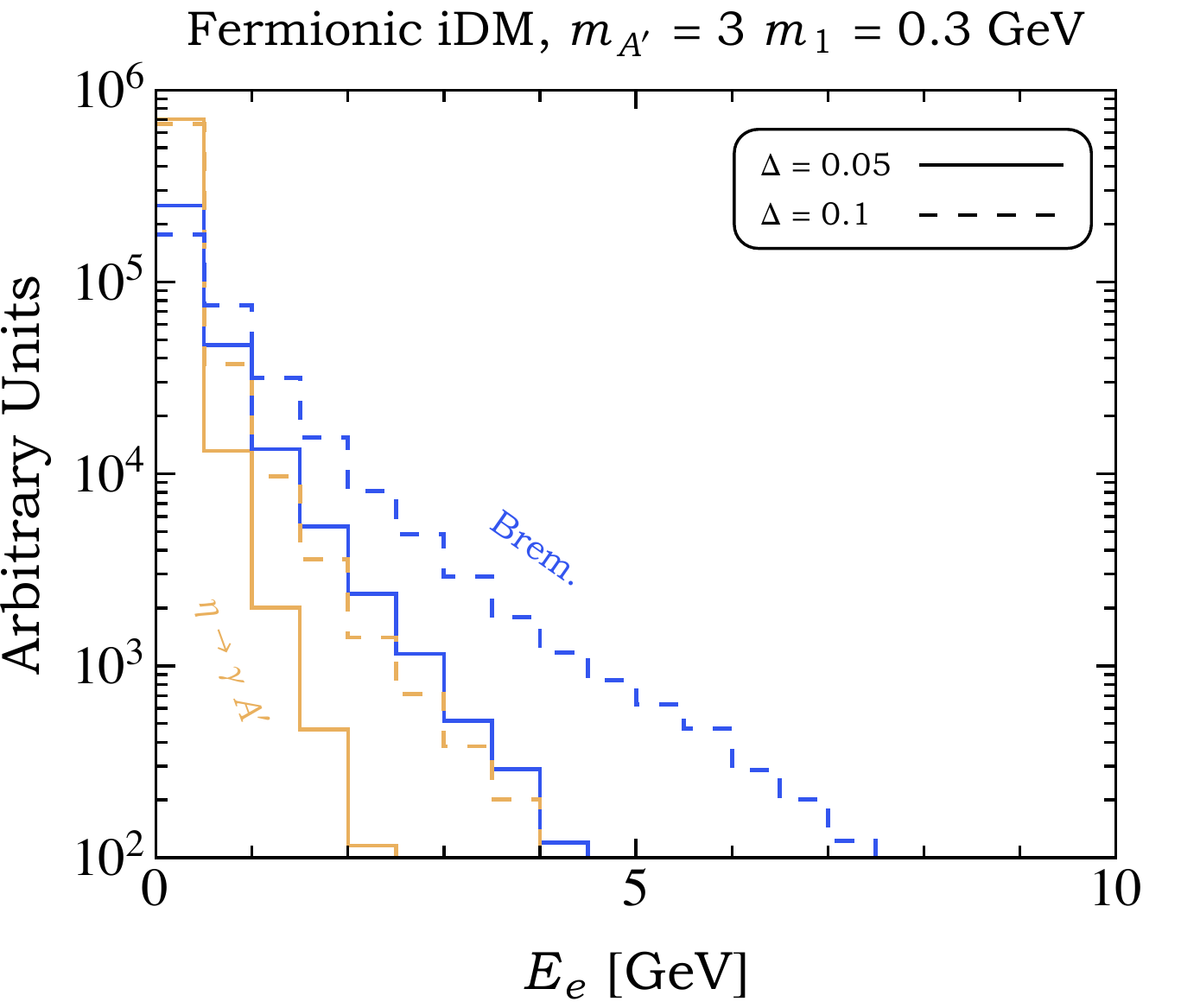}
\includegraphics[width=0.49\textwidth]{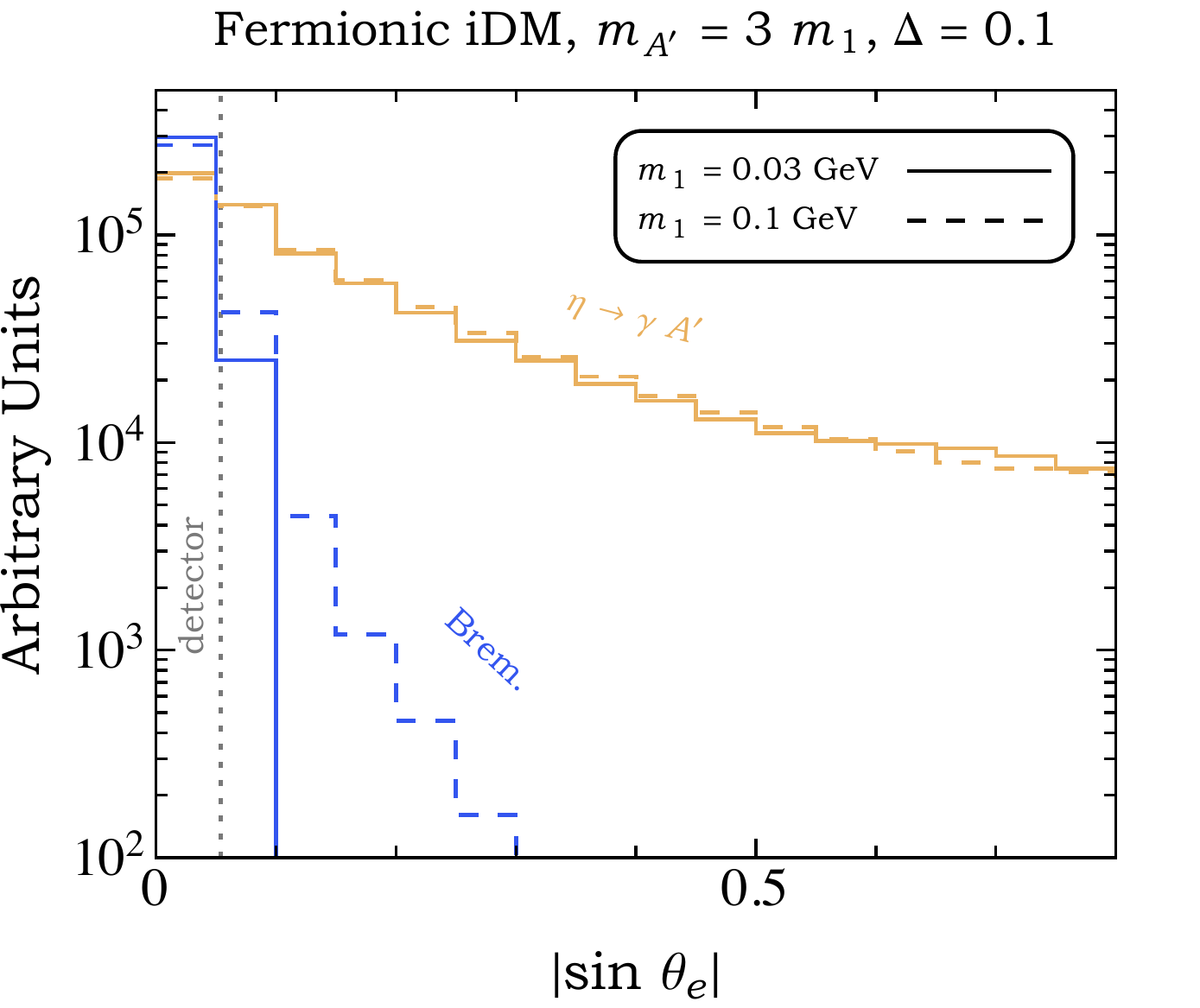}
\caption{Signal kinematics of $\Ap \to \chi_1 \chi_2 \to \chi_1 \chi_1 e^+ e^-$ for dark photons produced from exotic eta meson decays (orange) and proton Bremsstrahlung (blue).  The left (right) panel displays energy (angular) distributions for electrons originating from $\chi_2$ decays before traveling through KMAG for $\mAp = 3 \, m_1$. In the left panel, we fix $m_1 = 0.1 \GeV$ and the solid (dashed) line corresponds to $\Delta = 0.05$ $(0.1)$. In the right panel, we fix $\Delta = 0.1$ and the solid (dashed) line corresponds to $m_1 = 0.03 \GeV$ ($0.1 \GeV$). The vertical gray dotted line in the right panel denotes the angular scale of the SeaQuest spectrometer.}
\label{fig:KiniDM}
\end{figure*}

In Fig.~\ref{fig:iDMthermal}, we fix $\epsilon$ in the $m_1 - \alpha_D$ plane such that $\chi_1$ freezes out with an adequate relic abundance. This figure therefore highlights the remaining viable iDM parameter space, assuming that $\chi_1$ constitutes the DM of the universe.  Larger $\alpha_D$ facilitates DM freeze-out for smaller values of $\epsilon$ (see Eq.~(\ref{eq:thermalm1})), suppressing the dark photon production rate at terrestrial experiments and alleviating existing constraints. The majority of such constraints are shown as gray shaded regions. However, in Fig.~\ref{fig:iDMthermal}, we have refrained from displaying the reach from searches for visible $\chi_2$ decays at the E137 experiment. As discussed above, these limits are sensitive to the particular choice of the ECAL energy deposition threshold, and, hence, do not robustly exclude any of the parameter space shown. The currently viable parameter space for thermal iDM involves values of the $U(1)_D$ coupling, $\alpha_D$, larger than  $\order{10^{-2}}$.  This is also arguably the most theoretically motivated parameter space, where $\alpha_D$ is comparable in strength to SM gauge couplings.

\subsection{Inelastic Dark Matter at SeaQuest}

In this section, we analyze the prospects for detecting signals of iDM through measurements of displaced electron pairs with the SeaQuest spectrometer. If $\mAp > m_1 + m_2$, then dark photons that are produced from the collision of the proton beam with the iron beam dump promptly decay to DM, $\chi_1$, and its excited state, $\chi_2$, i.e., $\Ap \to \chi_1 \chi_2$. $\chi_2$ is naturally long-lived and visibly decays to SM electrons after traversing the iron shield, $\chi_2 \to \chi_1 e^+ e^-$. Although the displaced electrons in this scenario do not reconstruct a resonance (as considered in Sec.~\ref{sec:SeaQuestAprime}) and do not point back to the primary interaction point, such decays still constitute an interesting avenue for detection at SeaQuest. In the discussion below, we assume that the displaced vertex is sufficient to reject SM background processes even without mass reconstruction of the electron pair or a pointing requirement.

As discussed in Sec.~\ref{sec:darkphoton}, we model the flux of dark photons produced from exotic meson decays, Bremsstrahlung, and Drell-Yan. 
The efficiency and total number of signal events are calculated using Eqs.~(\ref{eq:TotalEff}) and (\ref{eq:Nsignal}). 
We demand that $\chi_2$ decays inside one of the three fiducial regions discussed above: $5 \m - 6 \m$, $5 \m - 9 \m$, or $5 \m - 12 \m$. 
The electron pair is required to satisfy the same kinematic requirements as explained below Eq.~(\ref{eq:TotalEff}). The signal is therefore similar to that considered in Sec.~\ref{sec:SeaQuestAprime}. However, for $\Delta \ll 1$, $\chi_1$ carries away a large fraction of the $\chi_2$ energy, implying that the electrons from the 3-body decay are significantly softer than those from visibly decaying dark photons. This is seen in Fig.~\ref{fig:KiniDM}, which shows the energy and angular distributions of electrons originating from $\chi_2$ decays.
As a result, the total efficiency is comparatively suppressed.

\begin{figure*}[t]
\hspace{-0.5cm}
\includegraphics[width=0.5\textwidth]{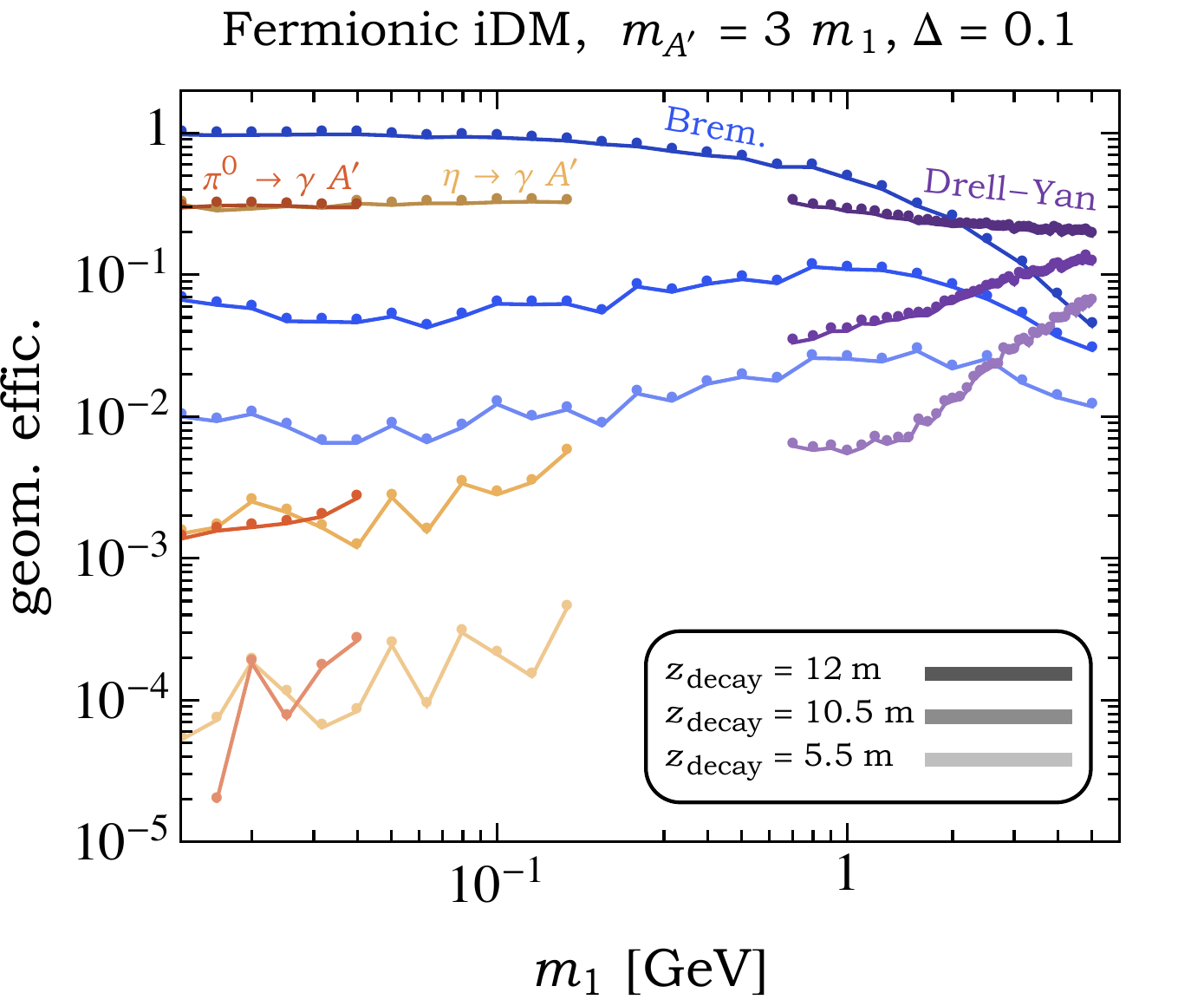} \hspace{-0.5cm}
\includegraphics[width=0.5\textwidth]{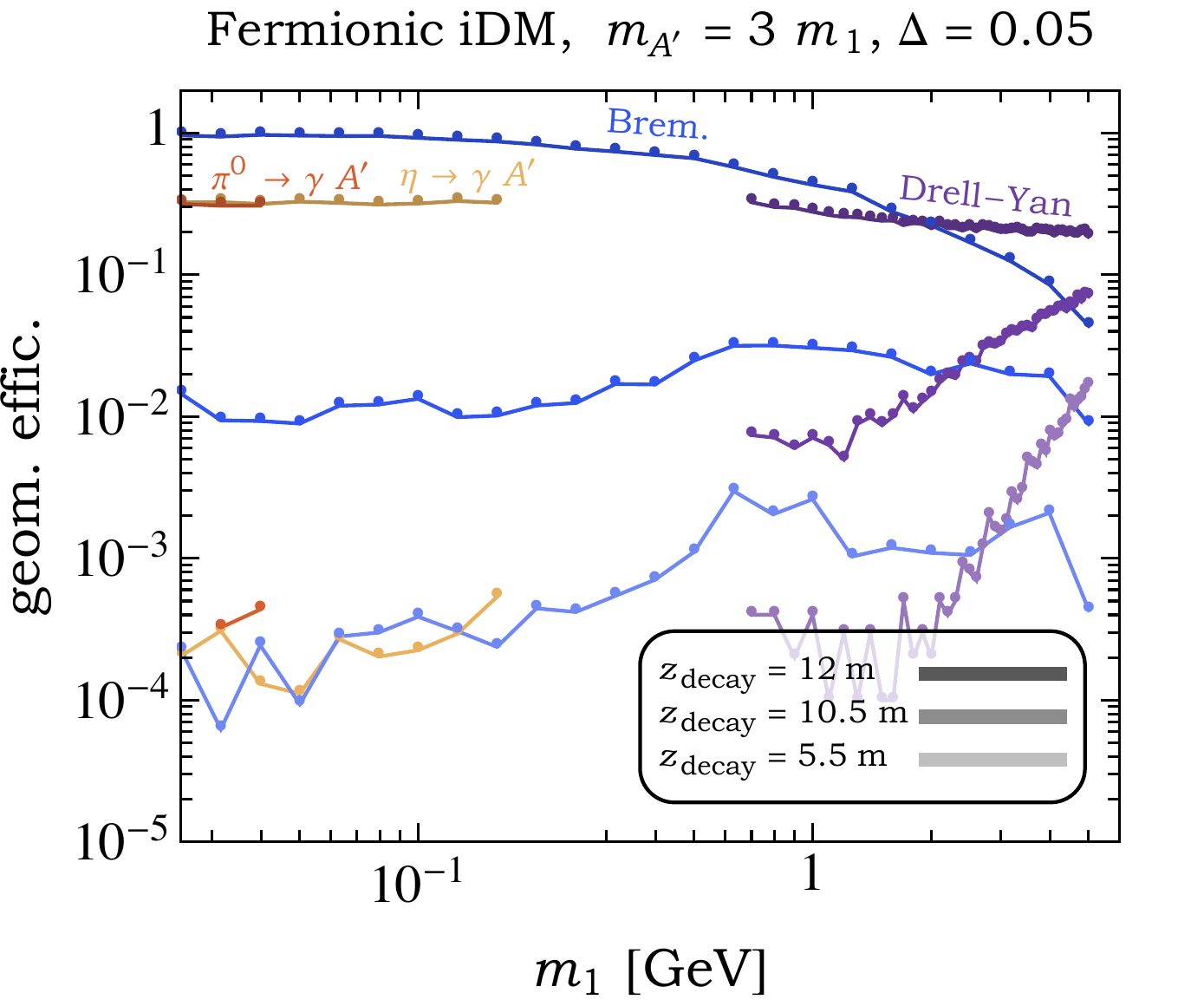} \hspace{-0.5cm}\\
\caption{In models of inelastic dark matter, the geometric efficiency for dark photons produced from proton Bremsstrahlung (blue), the decays of pions (red) or eta mesons (orange), and Drell-Yan (purple). For each of these production channels, we assume that $\chi_2$ decays to an electron pair after traveling $5.5 \m$ (bottom line), $10.5 \m$ (middle line), or $12 \m$ (top line). For these latter two decay points, the electron pair only traverses a fraction of the KMAG magnet.}
\label{fig:iDMeffic}
\end{figure*}

In Fig. \ref{fig:iDMeffic}, we show the fraction of signal events that pass the geometric selection for the dominant production channels as a function of $m_1$. As in Fig.~\ref{fig:efficiencyDarkPhoton}, we fix the position at which $\chi_2$ decays to an electron pair to the representative lengths: $5.5 \m$ (bottom line), $10.5 \m$ (middle line), and $12 \m$ (top line). The overall behavior of these acceptances as a function of $m_1$ follows from the same discussion that was previously given for Fig.~\ref{fig:efficiencyDarkPhoton}. However, compared to the minimal dark photon scenario ($\Ap \to e^+ e^-$), the geometric efficiency in models of iDM is extremely sensitive to the fraction of KMAG that the soft electron pairs must travel through. In Fig.~\ref{fig:iDMeffic}, this is most noticeable for dark photons produced via meson decays, as the electrons are peaked towards smaller energies. Indeed, for the mass splitting $\Delta = 0.05$, the energies of the decay products are significantly suppressed, and we find no electron pairs originating from pion or eta decays that are able to penetrate the full length of KMAG (for $z_{\rm{decay}}=5.5 \m$). 

\begin{figure}[t] 
\vspace{0.cm}
\includegraphics[width=0.497\textwidth]{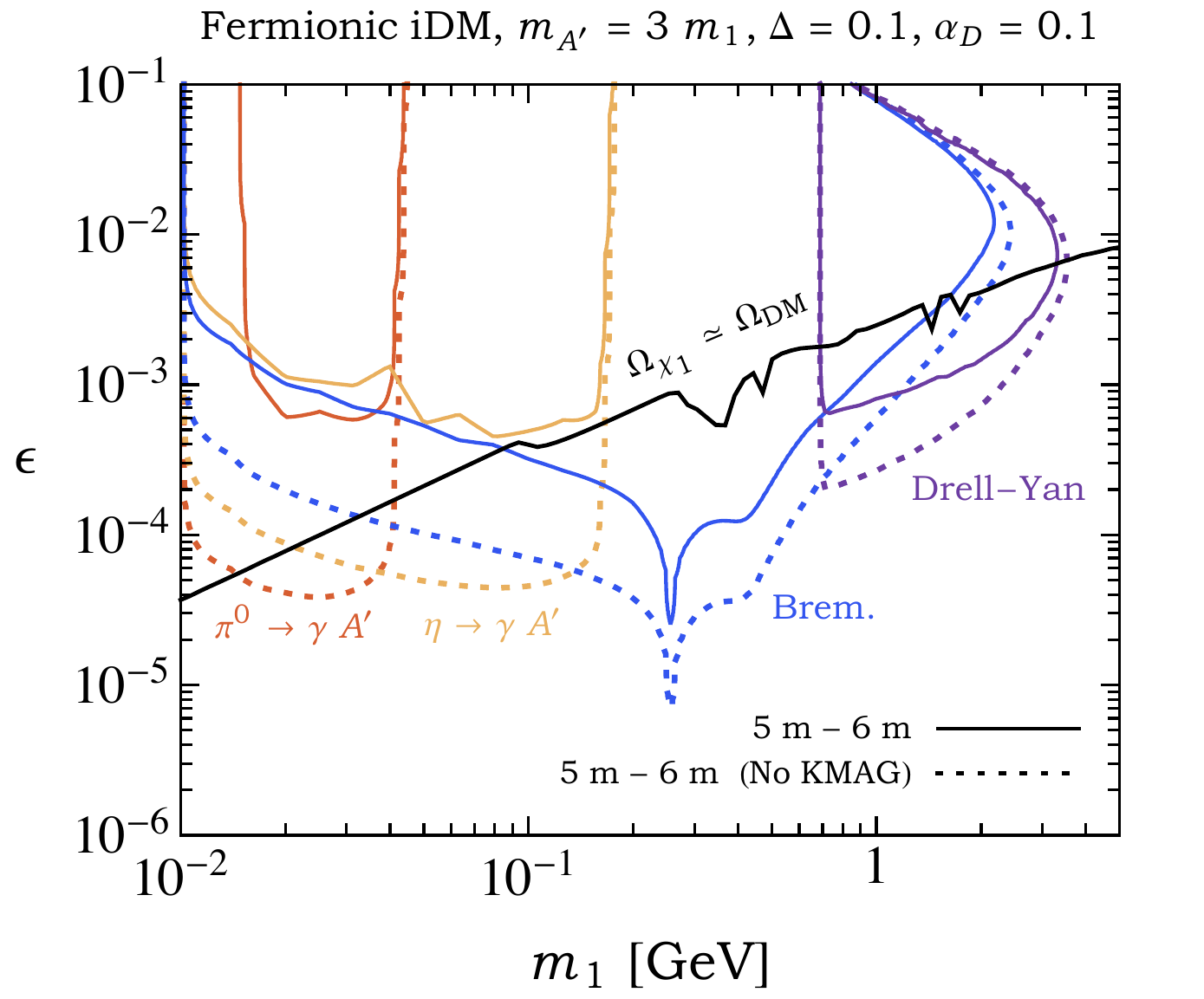}
\caption{The projected Phase I SeaQuest sensitivity to the inelastic dark matter parameter space using the $5 \m - 6 \m$ fiducial decay region for $\mAp / m_1 = 3$, $\Delta = 0.1$, and $\alpha_D = 0.1$. The various contours correspond to 10 dielectron signal events, for dark photons produced from meson ($\pi^0,\eta$) decays, proton Bremsstrahlung, and Drell-Yan. The solid (dotted) contours correspond to a setup with (without) the magnetic field of KMAG. Along the black contour, the abundance of $\chi_1$ agrees with the observed dark matter energy density.}
\label{fig:iDM_Contributions}
\end{figure}

In Figs.~\ref{fig:iDMSummary} and \ref{fig:iDMthermal}, the solid (dashed or dotted) purple contours enclose regions of parameter space in which more than 10 displaced signal events are expected at Phase I (Phase II) of SeaQuest. We adopt the minimal fiducial decay region of $5 \m - 6 \m$ for Phase I, while for Phase II, moving from the darker to lighter dashed purple contours corresponds to the three fiducial decay regions: $5 \m - 6 \m$, $5 \m - 9 \m$, and $5 \m - 12\m$. We note that for $\Delta = 0.05$ in the right panel of Fig.~\ref{fig:iDMthermal}, our simulation suffers from limited statistics, which results in reduced geometric efficiencies and non-physical jagged features in the projected SeaQuest sensitivities. In dotted purple, we also show the Phase II reach for the $5 \m - 6 \m$ decay region, assuming that the electrons do not have to travel through the magnetic field of KMAG. Such a setup is feasible assuming that triggering on the displaced vertex 
is sufficient to reject background. Furthermore, the geometric efficiency of events originating from meson decays is greatly enhanced without the strong magnetic field of KMAG. This is illustrated in Fig.~\ref{fig:iDM_Contributions}, which shows the various contributions from the dominant $\Ap$ production mechanisms for a Phase I setup with or without the magnetic field of KMAG. As in Sec.~\ref{sec:SeaQuestAprime}, we have assumed that such a search for energetic displaced electrons would be nearly background-free. Detecting such visible signals is potentially feasible for sufficiently large mass splitting, $\Delta \gtrsim \order{10^{-2}}$.

The sensitivity of SeaQuest is complementary to other future and proposed experiments.  Two of these, Belle-II (discussed above) and the proposed Light Dark Matter eXperiment (LDMX) at SLAC are also shown in Figs.~\ref{fig:iDMSummary} and \ref{fig:iDMthermal}.  LDMX aims to detect DM particle production by measuring the incident momenta of individual electrons impinging on a thin tungsten target and the full final state of these interactions. DM production events are characterized by dramatic energy loss and significant transverse momentum transfer in the target, with no other visible particles carrying away this energy.  This signature is relevant to iDM models so long as $\chi_2$ decays downstream of the calorimeter.  The expected sensitivity of LDMX, assuming 
 10-event sensitivity with $\order{10^{16}}$ EOT using the 4 or 8 GeV LCLS-2 electron beam~\cite{Izaguirre:2014bca}, is shown in green in Figs.~\ref{fig:iDMSummary} and \ref{fig:iDMthermal}.  Visible signals of $\chi_2\rightarrow \chi_1 e^+e^-$ with decays deep inside the LDMX calorimeter may also be detectable, but we leave a detailed investigation of this scenario to future work.  
 
 We also note that the same displaced visible signals discussed here can be observed at low-energy $e^+ e^-$ colliders such as BaBar and Belle-II, as well as at the LHC~\cite{Izaguirre:2015zva}. Although not explicitly shown in Figs.~\ref{fig:iDMSummary} and \ref{fig:iDMthermal}, these experiments are expected to have complementary sensitivity and probe viable parameter space for $m_1 \gtrsim \text{several} \times \text{GeV}$ and $\Delta \gtrsim 0.1$. Future searches at MiniBooNE and the proposed BDX experiment may also be sensitive to viable regions of parameter space for $m_1 \lesssim \text{GeV}$~\cite{Izaguirre:2017bqb}.

Several parameters were held fixed in Figs.~\ref{fig:iDMSummary} and \ref{fig:iDMthermal}, e.g., $\mAp=3 \, m_1$. 
The signals we have considered are suppressed if $\mAp < m_1 + m_2$ , although production of $\chi_1$ and $\chi_2$ is still possible through an off-shell dark photon. 
Additional probes relevant to this scenario include searches for direct decays to the SM, $\Ap \to \ell^+ \ell^-$, as in Sec.~\ref{sec:SeaQuestAprime}. For $\mAp / m_1 \gg 1$, $\Ap$ production is suppressed or kinematically inaccessible at low-energy accelerators and the viability of the signals discussed throughout this work is significantly weakened. Higher energy searches, such as the proposed FASER experiment~\cite{Feng:2017uoz}, offer an obvious advantage for such models. 
Furthermore, the cosmology and phenomenology of iDM are both sensitive to the $\chi_2 - \chi_1$ mass splitting, $\Delta$ (in Figs.~\ref{fig:iDMSummary} and \ref{fig:iDMthermal} we fix $\Delta=0.1$ and 0.05). As discussed in Sec.~\ref{eq:iDMmodel}, the cosmologically favored DM mass is exponentially sensitive to $\Delta$ (see Eq.~(\ref{eq:thermalm1})). In particular, for $\Delta > 0.1$, larger values of $\epsilon$ are required for fixed hidden sector masses, and less experimentally viable parameter space is currently accessible. 

For $\Delta \lesssim \order{10^{-2}}$, the electrons from visible $\chi_2$ decays are easily swept outside of the spectrometer by the KMAG magnet,  decreasing the projected sensitivity of SeaQuest to negligible levels. Decay searches such as the SeaQuest one discussed here become ineffective for such small splittings, while (for $\Delta \gtrsim \order{10^{-6}}$) direct detection of $\chi_1 X\rightarrow \chi_2 X$ (where $X$ is an electon or nucleon) remains kinematically suppressed. In contrast, high-intensity experiments searching for invisible signals (as in LDMX or Belle-II) or DM scattering play a crucial role in testing this possibility.
Furthermore, if $\Delta \lesssim (m_1 / \text{MeV} )^{-1}$, $\chi_2 \to \chi_1 e^+ e^-$ is kinematically forbidden.  This qualitatively alters the cosmological phenomenology: although scattering processes still deplete the cosmological population of $\chi_2$ at late times, several experiments can constrain even this small residual abundance.  In particular, parts of the parameter space are constrained by considerations of CMB energy injection from $\chi_1-\chi_2$ coannihilation and/or $\chi_2$ decay, while other regions are constrained by $\chi_2$ downscattering in low-threshold direct detection experiments such as CRESST II~\cite{Toro:2018}.  

%

\section{Leptophilic Higgs Bosons}
\label{sec:leptoHiggs}

It has long been appreciated that new light physics that is feebly coupled to the SM can potentially resolve the $\sim 3.5 \sigma$ discrepancy between the observed and predicted value of the anomalous magnetic moment of the muon, $(g-2)_\mu$~\cite{Gninenko:2001hx,Fayet:2007ua,Pospelov:2008zw,Bennett:2006fi}. One such example is the minimal dark photon, as discussed in Sec.~\ref{sec:SeaQuestAprime}, which has recently been excluded as a viable explanation~\cite{Batley:2015lha}. However, there are possible alternatives that can alleviate the tension between theory and experiment, such as models of $L_\mu - L_\tau$ gauge bosons or light scalars that couple predominantly to leptons~\cite{Altmannshofer:2014pba,Altmannshofer:2016brv,Kinoshita:1990aj,Chen:2015vqy,Batell:2016ove}. In this section, we briefly discuss SeaQuest's capability to probe this latter scenario. 

\begin{figure}[t]
\hspace{-0.5cm}
\includegraphics[width=0.5\textwidth]{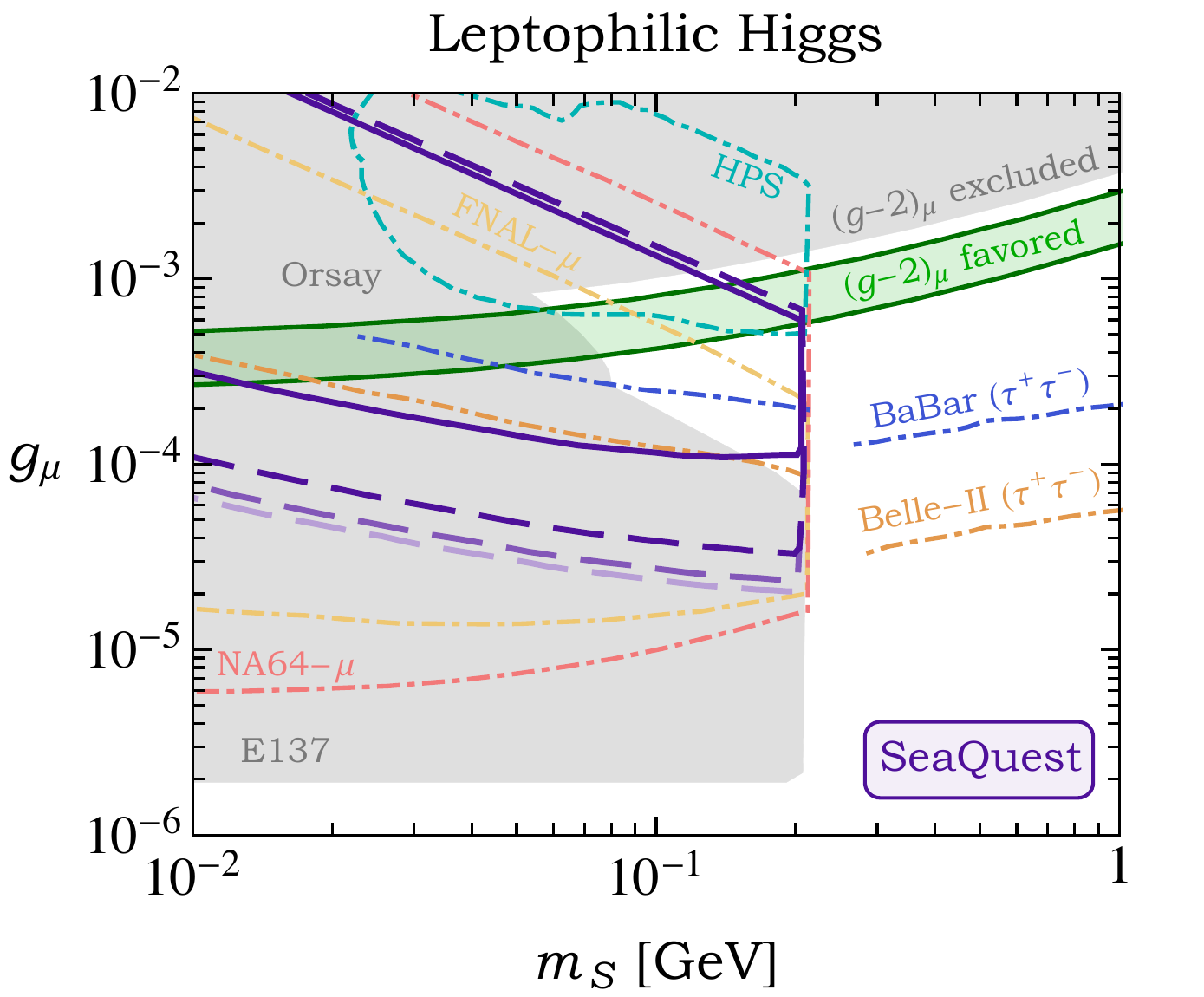} \hspace{-0.5cm}
\caption{Existing constraints (shaded gray) and projected sensitivities (color) to models of leptophilic Higgs bosons, $S$. The shaded regions are excluded by E137~\cite{Bjorken:1988as}, Orsay~\cite{Davier:1989wz}, and measurements of the anomalous magnetic moment of the muon~\cite{Grange:2015fou,Mibe:2010zz}. The colored lines correspond to the projected reach of HPS (cyan)~\cite{Ilten:2015hya,Ilten:2016tkc}, the proposed muon fixed-target NA64-like (pink) and Fermilab (yellow)~\cite{Chen:2017awl} experiments, and proposed searches at BaBar (blue), Belle-II (orange)~\cite{Batell:2016ove}, and SeaQuest (purple), as described in the text. The projected reach of SeaQuest is shown as in Fig.~\ref{fig:visAp}. For Phase I (solid purple), we conservatively fix the fiducial decay region to $5 \m - 6 \m$. For Phase II (dashed purple), moving from darker to lighter contours corresponds to the fiducial decay regions: $5 \m - 6 \m$, $5 \m - 9 \m$, and $5 \m - 12 \m$, respectively. Also shown is the region of couplings and masses that alleviate the tension between the observed and predicted values of the anomalous magnetic moment of the muon (green).}
\label{fig:muon}
\end{figure}

We investigate a simplified low-energy model in which a light scalar, denoted by $S$, couples universally to SM leptons proportional to their mass, i.e.,
\be
- \mathcal{L} \supset \sum\limits_{\ell = e, \mu, \tau} \frac{m_\ell}{\Lambda} ~ S ~ \bar{\ell} \ell
~,
\ee
where $\Lambda$ is the scale associated with the new physics that generates these couplings.  
Since we are interested in this model's ability to resolve the $(g-2)_\mu$ discrepancy, we normalize the interactions to the muon coupling, $g_\mu$,
\be
\label{eq:darkHiggs}
- \mathcal{L} \supset g_\mu \sum\limits_{\ell = e, \mu, \tau} \frac{m_\ell}{m_\mu} ~ S ~ \bar{\ell} \ell
~.
\ee
For $m_S \sim 10 \MeV - 1 \GeV$, couplings of size $g_\mu \sim 10^{-4} - 10^{-3}$ resolve the disparity in $(g-2)_\mu$~\cite{Chen:2015vqy,Batell:2016ove}. Such interactions can be introduced in a gauge invariant manner through a dimension-5 operator, involving the SM Higgs doublet, $H$,
\be
\mathcal{O}_S \sim \frac{S}{\Lambda} ~ \bar{E}_L \, H \, e_R
~,
\ee
which can be UV-completed, e.g., in two-Higgs-doublet models involving additional singlet scalars or vector-like quarks~\cite{Chen:2015vqy,Batell:2016ove}. Hereafter, we adopt the simplified model of Eq.~(\ref{eq:darkHiggs}) in order to describe the relevant low-energy phenomenology. 

Refs.~\cite{Essig:2010gu, Batell:2016ove, Chen:2017awl} have proposed the construction of muon beam fixed-target experiments to explore the low-mass parameter space of lepton-coupled scalars. As discussed in Ref.~\cite{Batell:2016ove}, proton beam dumps may also offer sensitivity to these models. Proton-nucleus collisions at SeaQuest lead to a large multiplicity of final state SM particles, including pions as well as muons from $\pi \to \mu \, \nu$. What is often thought of as unwanted QCD ``baggage" is actually a new source of light exotic states if these muons radiate the scalar, $S$, in coherent muon-iron nucleus collisions. For $m_S \lesssim 2 \, m_\mu$, $S$ dominantly decays to $e^+ e^-$ with a proper lifetime that is macroscopic for values of $g_\mu$ that are motivated by $(g-2)_\mu$,
\be
\tau_S \sim \order{10} \cm \times \left(\frac{g_\mu}{10^{-4}}\right)^{-2} \, \left( \frac{m_S}{100 \MeV} \right)^{-1}
~.
\ee
For $m_S \sim 100 \MeV$, loop-induced decays to pairs of photons are also relevant, but are typically subdominant, i.e., $\text{BR}(S \to \gamma \gamma) \lesssim 10 \%$~\cite{Batell:2016ove, Chen:2017awl}. 

As in Sec.~\ref{sec:darkphoton}, we model pion production at SeaQuest using {\tt PYTHIA 8.2}~\cite{Sjostrand:2014zea}, incorporating decays to muons. For simplicity, we demand that these decays occur within the first pion collision length of iron (13.59 cm)~\cite{Patrignani:2016xqp}, in which case energy loss prior to decay can be ignored. We simulate muon Bremsstrahlung production of $S$ with {\tt MadGraph5}~\cite{Alwall:2011uj}, as described for electron Bremsstrahlung of dark photons in Sec.~\ref{sec:iDM}. As in the previous sections, we demand that $S$ decays to an electron pair that remains within the geometric acceptance of the spectrometer, incorporating the fiducial decay regions of $5 \m - 6 \m$, $5 \m - 9 \m$, and $5 \m - 12\m$. 

The efficiency and signal rate are calculated similar to Eqs.~(\ref{eq:TotalEff}) and (\ref{eq:Nsignal}). However, these expressions must be generalized since the high-energy muons from pion decays are not monochromatic or uniform in direction and can traverse a significant portion of FMAG before radiating the scalar, $S$. For $E_\mu \sim 1 \GeV- 100 \GeV$, muon energy loss dominantly occurs through ionization~\cite{Patrignani:2016xqp} as it traverses FMAG, which is described by the stopping power, $dE_\mu / dz_\mu$, where $z_\mu$ is the muon penetration length. We also model the bending of the muons from the magnetic field of FMAG as they travel through the iron dump. Analogous to Eq.~(\ref{eq:TotalEff}), the number of signal events from production and decay of $S$ is given by~\cite{Chen:2017awl}
\begin{widetext}
\begin{align}
\label{eq:NsignalMuon}
N_\text{signal} \simeq \text{BR} (S \to e^+ e^-) n_\text{atom} m_S \Gamma_S  \int d E_\mu^0 \frac{d N_\mu}{d E_\mu^0}  \int_{E_\mu^\text{min}}^{E_\mu^0} d E_\mu  \frac{\sigma_\text{Brem.} (E_\mu)}{|d E_\mu / d z_\mu|}  \int_{z_\text{min} - z_\mu}^{z_\text{max} - z_\mu}  d z \sum\limits_{\text{events $\in$ geom.}}^{E_\mu} \frac{e^{- z  (m_S / p_z^S)  \Gamma_S}}{N_\text{MC}   p_z^S}
.
\end{align}
\end{widetext}
As in Eq.~(\ref{eq:TotalEff}), $z_\text{min} - z_\text{max}$ defines the fiducial decay region as measured from the front of FMAG, $N_\text{MC}$ is the total number of simulated events (of fixed muon energy), and the sum is performed over only those events that remain within the geometry of the SeaQuest spectrometer. In Eq.~(\ref{eq:NsignalMuon}), we also integrate over $E_\mu^0$ and $E_\mu$, where $E_\mu^0$ is the initial energy of the muon from the decay of the pion, and $E_\mu$ is the muon energy after traversing some finite distance ($z_\mu$) in FMAG before radiating the scalar, $S$. $dN_\mu/dE_\mu^0$ is the total number of muons produced per initial energy bin within the first pion interaction length, $n_\text{atom} \simeq 8.5 \times 10^{22} \cm^{-3}$ is the number density of target iron nuclei, and $\sigma_\text{Brem.}$ is the muon Bremsstrahlung cross-section. 
Finally, $z_\mu^\text{min} \simeq 0 \m$ and $z_\mu^\text{max} = 5 \m$ denote the FMAG region of the beam line, which allows us to relate $z_\mu$ and $E_\mu$ as well as define $E_\mu^\text{min}$ through
\be
z_\mu (E_\mu, E_\mu^0) = z_\mu^\text{min} + \int_{E_\mu}^{E_\mu^0} \frac{d E_\mu^\prime}{|d E_\mu^\prime / d z_\mu|}
~,
\ee
 and
\be
z_\mu (E_\mu^\text{min}, E_\mu^0) = z_\mu^\text{max}
~.
\ee
In practice, soft muons are easily deflected by the magnetic field of FMAG, which sets the lower bound on $E_\mu$ in the integral of Eq.~(\ref{eq:NsignalMuon}).

The projected sensitivity of SeaQuest to the simplified model of Eq.~(\ref{eq:darkHiggs}) is shown in Fig.~\ref{fig:muon},
both at Phase I (solid purple) and Phase II (dashed purple) in the $m_S - g_\mu$ plane, demanding 10 dielectron signal events and assuming negligible SM background. For each point in parameter space, the coupling to electrons is given by $g_\mu \times (m_e / m_\mu)$ (see Eq.~(\ref{eq:darkHiggs})). As in the previous sections, we adopt the minimal  $5 \m - 6 \m$ fiducial decay region for Phase I, while for Phase II, moving from the darker to lighter dashed contours corresponds to decay regions of $5 \m - 6 \m$, $5 \m -  9 \m$, and $5 \m - 12 \m$, respectively. The reach is strongly diminished for $m_S \gtrsim 2 \, m_\mu$ since the $S$ is shorter-lived and the branching fraction to electrons is suppressed by a factor of $\order{10^4}$. 

Also shown are existing constraints and the projected reach of other proposed experiments. 
Electron beam dump experiments such as E137~\cite{Bjorken:1988as} (see Sec.~\ref{sec:iDMconstraints}) and Orsay~\cite{Davier:1989wz} are sensitive to energy deposition from $S \to e^+ e^-$, provided that $S$ is sufficiently long-lived~\cite{Batell:2016ove,Chen:2017awl}. These exclusions are shown as shaded gray regions in Fig.~\ref{fig:muon}.
Dedicated muon fixed-target NA64-like~\cite{Gninenko:2014pea,Gninenko:2018tlp} and Fermilab experiments were recently proposed in Ref.~\cite{Chen:2017awl} and the corresponding reach is shown as the pink and yellow contours in Fig.~\ref{fig:muon}, respectively, assuming 100\% signal efficiency. Searches at the $B$-factories BaBar and Belle-II for $\tau^+ \tau^-$ associated production of $S \to \ell^+ \ell^-$ have also been proposed in Ref.~\cite{Batell:2016ove}. These searches are sensitive to the parameter space above the blue and orange lines, respectively. We also take from Ref.~\cite{Batell:2016ove} the projected sensitivity of the HPS experiment~\cite{Ilten:2015hya,Ilten:2016tkc} to the displaced decay, $S \to e^+ e^-$, which is shown in cyan.
The region of couplings and masses that alleviate the tension between the observed and predicted values of the anomalous magnetic moment of the muon are highlighted in green, while values of $g_\mu$ that are in conflict with such measurements are shown as shaded gray. Emission of $S$ in rare $B$-meson decays, e.g., $B \to K \, S \to K \, \mu^+ \mu^-$, can lead to deviations in the muon spectra as measured at LHCb~\cite{Aaij:2015tna,Batell:2016ove}. Although such measurements are potentially sensitive to large regions of parameter space, the corresponding bounds are not explicitly shown in Fig.~\ref{fig:muon} since they depend on the specific UV-completion of the toy model in Eq.~(\ref{eq:darkHiggs}), i.e., on the couplings of $S$ with quarks.

Fig.~\ref{fig:muon} illustrates that an ECAL upgrade to SeaQuest would allow for significant discovery potential for new light scalars motivated by the $(g-2)_\mu$ anomaly. The Fermilab muon fixed-target experiment proposed in Ref.~\cite{Chen:2017awl} consists of utilizing the 3 GeV muon beam for the ongoing $(g-2)_\mu$ experiment~\cite{Chapelain:2017syu}. In comparison, a significant number of muons produced at SeaQuest are of much larger energy, and the enhanced boost of $S$ allows a displaced electron search to be sensitive to larger values of $g_\mu$ (smaller proper lifetimes) that are capable of explaining the enduring $(g-2)_\mu$ discrepancy.

\section{Beyond the Electron Channel}
\label{sec:other}

In this section, we briefly discuss the prospects for SeaQuest to detect other long-lived particles, such as minimal dark Higgs bosons and axion-like particles (ALPs). Contrary to the previous sections, we consider searches for final states other than electron pairs, such as SM photons and pions. Such final states potentially introduce complications from additional SM background processes compared to searches for displaced electrons. A detailed investigation of the viability of these other channels is beyond the scope of this work. When relevant, we will instead discuss minor modifications to the existing spectrometer in order to properly suppress SM backgrounds. Hence, in evaluating the projected reach of such SeaQuest-like setups, we optimistically assume that background processes are negligible. It would be very interesting to have dedicated feasibility studies performed by the SeaQuest collaboration.

\subsection{Minimal Dark Higgs}

Models of extended Higgs sectors often involve new scalar singlets that directly mix with the SM Higgs, which is controlled by the mixing angle, $\sin{\theta}$. In this case, the scalar singlet ($\p$) possesses feeble couplings with SM fermions ($f$) proportional to $\sin{\theta}$. We parametrize the relevant interactions of the low-energy theory as
\be
\label{eq:mindarkHiggs}
- \mathcal{L} \supset \sin{\theta} ~ \frac{m_f}{v} ~ \p \, \bar{f} f
~,
\ee
where $v = 246 \GeV$ is the SM Higgs VEV. For $m_\p \lesssim 200 \MeV$, the dominant decay mode is to pairs of SM electrons. In the limit that $m_e \ll m_\p$, the corresponding partial width is approximately
\be
\Gamma (\p \to e^+ e^-) \simeq \frac{m_e^2 \, \sin^2{\theta}}{8 \pi \, v^2} \, m_\p
~.
\ee
For $m_\p \gtrsim 200 \MeV$, $\p$ dominantly decays to muons and hadrons. For decays to hadrons, there is large disagreement  in the literature among various computations of the scalar width  that differ up to several orders of magnitude for certain values of $m_\p$~\cite{Gunion:1989we,Voloshin:1985tc,Grinstein:1988yu,Raby:1988qf,Donoghue:1990xh,Truong:1989my}. This is shown explicitly, e.g., in Fig.~1 of Ref.~\cite{Clarke:2013aya}. In computing SeaQuest's sensitivity to dark Higgs bosons, we will vary the width of $\p$, $\Gamma_\p$, within the total range as computed in Refs.~\cite{Gunion:1989we,Voloshin:1985tc,Grinstein:1988yu,Raby:1988qf,Donoghue:1990xh,Truong:1989my}. 

At proton beam dumps, the majority of dark Higgs bosons are produced from the decays of heavy mesons, such as $K \to \pi \p$ and $B \to K \p$. For a more detailed discussion of dark Higgs production at proton fixed-target experiments, see, e.g., Refs.~\cite{Feng:2017vli,Alekhin:2015byh,Gligorov:2017nwh,Evans:2017lvd,Schmidt-Hoberg:2013hba,Batell:2009jf}. Previous beam dump experiments strongly constrain dark Higgs bosons for $m_\p \lesssim \text{few} \times \order{100} \MeV$. Hence, we will not compute contributions to $\p$ production from $K \to \pi \p$ since this is kinematically suppressed for $m_\p \gtrsim m_K - m_\pi \sim 350 \MeV$. As in the previous sections, we model the production of $B$-mesons using {\tt PYTHIA 8.2}~\cite{Sjostrand:2014zea} and manually decay these mesons to dark Higgs final states. 

\begin{figure}[t]
\hspace{-0.5cm}
\includegraphics[width=0.5\textwidth]{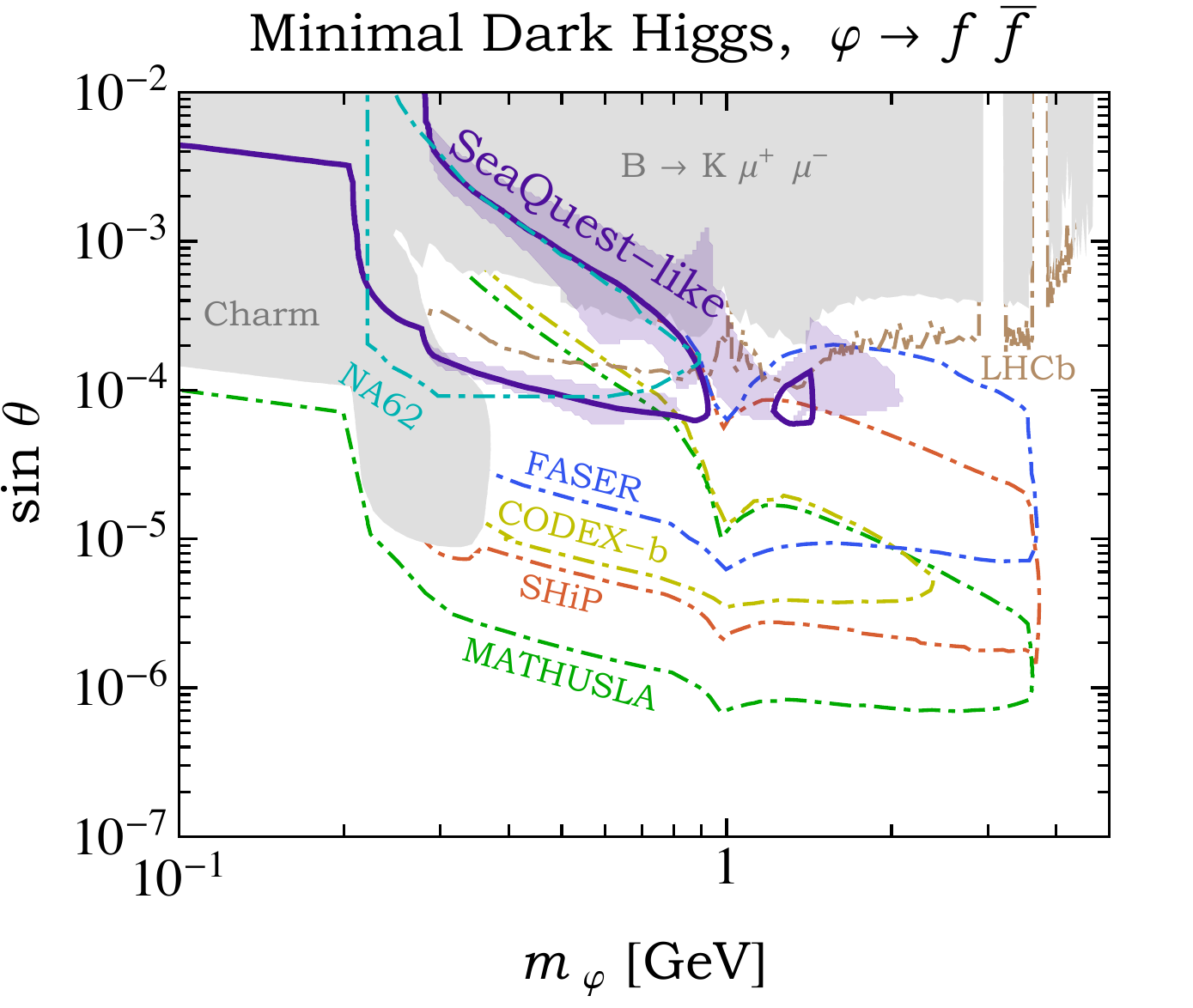} \hspace{-0.5cm}
\caption{Sensitivity of a SeaQuest-like experiment (see text) at Phase II (corresponding to 10 signal events) to displaced decays of minimal dark Higgs bosons into pairs of SM leptons and hadrons. We fix the fiducial decay region to $7 \m - 12 \m$, assuming that a few meters of additional iron is added to FMAG. The dark purple contour corresponds to the projected SeaQuest reach when the $\p$ width, $\Gamma_\p$, is fixed to the calculations of Refs.~\cite{Donoghue:1990xh} and \cite{Gunion:1989we}. Instead, in the light shaded purple region, we vary $\Gamma_\p$ within the full range as calculated in Refs.~\cite{Gunion:1989we,Voloshin:1985tc,Grinstein:1988yu,Raby:1988qf,Donoghue:1990xh,Truong:1989my}.
The gray region denotes the parameter space that is already excluded by past experiments~\cite{Evans:2017lvd} (Charm~\cite{Bergsma:1985qz} and LHCb measurements of $B\to K^{(*)}\mu^+\mu^-$~\cite{Aaij:2016qsm}). Also shown are the projected reach of a beam dump run of NA62 (cyan)~\cite{na62cern}, a 300 $\text{fb}^{-1}$ run at LHCb (brown)~\cite{Aaij:2016qsm,Gligorov:2017nwh}, and the proposed SHiP (red)~\cite{Alekhin:2015byh}, FASER (blue)~\cite{Feng:2017vli}, CODEX-b (yellow)~\cite{Gligorov:2017nwh}, and MATHUSLA (green)~\cite{Evans:2017lvd,Chou:2016lxi} experiments.}
\label{fig:darkHiggs}
\end{figure}

We utilize Eqs.~(\ref{eq:TotalEff}) and (\ref{eq:Nsignal}) in computing the projected reach of SeaQuest for the model of Eq.~(\ref{eq:mindarkHiggs}). However, since $\p$ decays predominantly to hadrons for $m_\p \gtrsim 100 \MeV$, we will focus on an inclusive search for displaced decays of $\p$, optimistically assuming negligible background for final states including SM pions and kaons. Our results are shown in Fig.~\ref{fig:darkHiggs} for an inclusive search for displaced vertices including decays to charged hadrons. To mitigate backgrounds from $K_L^0 \to 2 \pi, 3 \pi$, we assume that an additional few meters of iron is placed upstream of FMAG, which could suppress such processes to negligible levels. In estimating the reach of this modified SeaQuest-like experiment, we demand that $\p$ decays within the fiducial region of $7 \m - 12 \m$. In Fig.~\ref{fig:darkHiggs}, the projected Phase II reach of SeaQuest corresponds to regions of parameter space in which 10 signal events are expected.

As mentioned above, the predicted $\p$ width, $\Gamma_\p$, varies within several orders of magnitude for $m_\p \sim \text{GeV}$. Most existing studies adopt a single estimate for $\Gamma_\p$ in calculating the reach of proposed experiments and then compare to other experimental projections that typically utilize different calculations of $\Gamma_\p$. As a result, various choices for $\Gamma_\p$ among different studies can misleadingly overestimate or underestimate the projected sensitivity of one experiment compared to another. In order to illustrate the effect of this uncertainty, we show SeaQuest's projected sensitivity in Fig.~\ref{fig:darkHiggs}, fixing $\Gamma_\p$ to the calculations of Refs.~\cite{Donoghue:1990xh} and \cite{Gunion:1989we} for smaller and larger masses, respectively (solid purple), and also vary $\Gamma_\p$ within the region defined by the minimum and maximum prediction for each value of $m_\p$ from Refs.~\cite{Gunion:1989we,Voloshin:1985tc,Grinstein:1988yu,Raby:1988qf,Donoghue:1990xh,Truong:1989my} (shaded purple). We then compare the sensitivity of SeaQuest to existing constraints and other upcoming and proposed searches. There exist similar uncertainties in the exclusions and projections of past and proposed experiments but are not explicitly shown in Fig.~\ref{fig:darkHiggs}.

SeaQuest's ability to produce $B$-mesons is significantly stifled by its smaller center of mass energy compared to higher-energy proton beam experiments such as NA62~\cite{na62cern}, SHiP~\cite{Alekhin:2015byh}, FASER~\cite{Feng:2017vli}, CODEX-b~\cite{Gligorov:2017nwh}, LHCb~\cite{Aaij:2016qsm}, and MATHUSLA~\cite{Evans:2017lvd,Chou:2016lxi}. As a result, SeaQuest's reach is correspondingly reduced. Regardless, for $m_\p \sim 0.5 \GeV$, an inclusive search for displaced $\p$ decays at Phase II would be sensitive to currently viable parameter space and larger coupling than longer baseline experiments such as SHiP and MATHUSLA. We note that for $m_\p \lesssim 1.5 \GeV$, our adopted central estimate for $\Gamma_\p$ is similar to those utilized in Refs.~\cite{Alekhin:2015byh,Feng:2017vli,Aaij:2016qsm,Aaij:2016qsm,Evans:2017lvd}, from which we have taken the projections for SHiP, MATHUSLA, and CODEX-b. 

\begin{figure}[t]
\hspace{-0.5cm}
\includegraphics[width=0.5\textwidth]{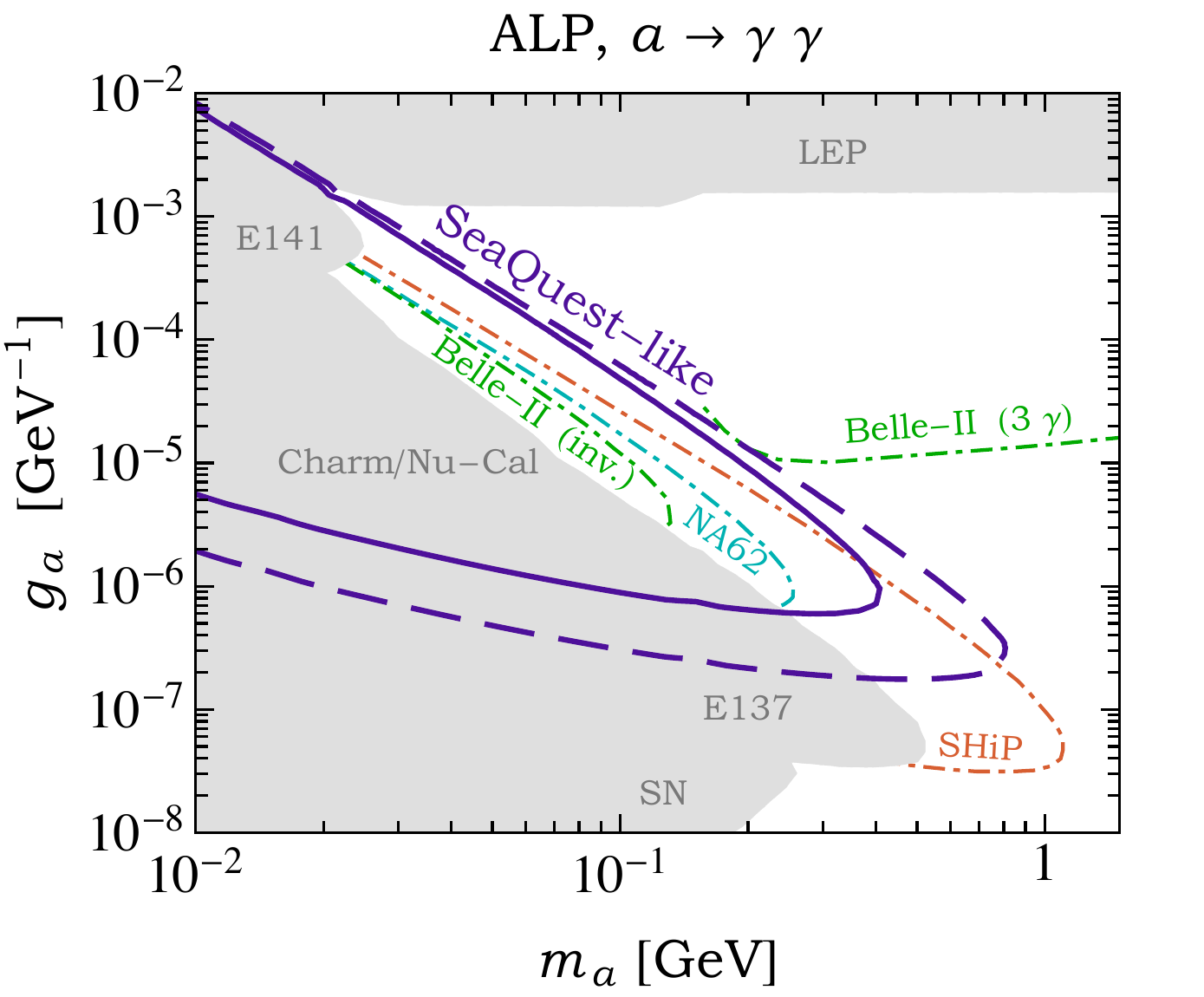} \hspace{-0.5cm}
\caption{Sensitivity to axion-like particles in the displaced diphoton channel of a Seaquest-like experiment (see text) at Phase I (solid purple) and Phase II (dashed purple), corresponding to 10 signal events. We conservatively fix the fiducial decay region to $7 \m - 8 \m$, assuming that a few meters of additional iron has been added behind FMAG. The gray region denotes the parameter space that is already excluded by past experiments~\cite{Dolan:2017osp}. Also shown are the projected reach of Belle-II (green)~\cite{Dolan:2017osp}, a beam dump run of NA62 (cyan)~\cite{na62cern}, and the proposed SHiP experiment (red)~\cite{Alekhin:2015byh}. Our estimated SeaQuest reach cannot be directly compared to the sensitivities of NA62 and SHiP, since the latter were calculated in Ref.~\cite{Dobrich:2015jyk} for axion-like particles produced from Primakoff reactions of the primary proton beam, whereas we have focused on production from the high-energy secondary photons from pion decays.}
\label{fig:ALP}
\end{figure}
%

\subsection{Axion-Like Particle}

We conclude this section with a brief discussion of SeaQuest's prospects for detecting electromagnetically coupled ALPs. In particular, we consider a minimal interaction, which couples a pseudoscalar ($a$) to the SM photon field strength, 
\be
\mathcal{L} \supset g_a \, a \, F_{\mu \nu} \, \tilde{F}^{\mu \nu}
~,
\ee
where $g_a$ has dimensions of inverse-mass. Such couplings often arise within the context of the QCD axion, where $a$ is identified as the pseudo-Goldstone boson associated with the spontaneous breaking of an additional global symmetry. In this case, the axion-photon coupling is naturally of size $g_a \sim \alpha_\text{em} / f_a$, where $f_a \gg \text{GeV}$ is the scale associated with the broken global symmetry. In our discussion, we treat $g_a$ as a free parameter of the low-energy theory. 

This interaction allows $a$ to decay to pairs of SM photons. The corresponding partial width is given by
\be
\Gamma (a \to \gamma \gamma) = \frac{g_a^2 \, m_a^3}{4 \pi}
~.
\ee
For $g_a \sim 10^{-7} \GeV^{-1} \times (m_a / \text{GeV})^{-3/2}$, the proper lifetime of $a$ is macroscopic ($\tau_a \sim \text{cm}$) and beam dump searches for displaced electromagnetic decays exclude large regions of parameter space for $m_a \lesssim 100 \MeV$ (see, e.g., Refs.~\cite{Jaeckel:2015jla,Dobrich:2015jyk,Dolan:2017osp} and references therein). As discussed below, proton-nucleus collisions at SeaQuest can lead to production of light ALPs, and searches for their decays can be conducted at SeaQuest after the proposed ECAL upgrade. 

Compared to the lepton signals of the previous sections, pointing and energy estimates from $a \to \gamma \gamma$ decays would only be possible through energy deposition in the ECAL. This decrease in pointing and energy resolution could potentially lead to significant background processes resulting from displaced decays of long-lived kaons, e.g., $K_L^0 \to \gamma \gamma$ and $K_L^0 \to 3 \pi^0 \to 6 \gamma$, similar to the discussion in Sec.~\ref{sec:analysis}. We estimate that an additional few meters of iron placed behind FMAG could suppress such processes to negligible levels, which might be viable without the need for tracking station 1. In estimating the reach of this modified SeaQuest-like experiment, we demand that the ALP decays to a photon pair, $a \to \gamma \gamma$, within the fiducial region of $7 \m -8 \m$, assuming that background processes are suppressed by the placement of additional iron behind FMAG.

Various mechanisms contribute to the production of electromagnetically coupled ALPs at proton fixed-target experiments. Primary Primakoff production from the fusion of two photons (one from the proton beam and the other from the target nucleus) is often considered as the leading process~\cite{Dobrich:2015jyk}. However, similar to the discussion in Sec.~\ref{sec:leptoHiggs}, we note that primary collisions of the proton beam with the iron dump lead to a plethora of high-energy secondary photons from the prompt decays of SM pions. A high-energy photon can then collide with an iron nucleus near the front of FMAG and transfer a significant portion of its energy to an ALP, i.e., $\gamma A \to a A$, where $A$ is an iron nucleus. Although direct production from the primary proton beam is enhanced by the ratio of the nuclear collision and radiation lengths in iron ($\sim 10$), we expect that production from secondary photons is a comparable or even dominant source of electromagnetically coupled ALPs since it is suppressed by one less power of $\alpha_\text{em}$. 

A full Monte Carlo analysis of $\pi^0 \to \gamma \gamma$ followed by $\gamma A \to a A$ is beyond the scope of this work. Instead, we estimate the signal yield as in Eq.~(17) of Ref.~\cite{Chen:2017awl}, 
\begin{widetext}
\begin{align}
\label{eq:NsignalALP}
N_\text{signal} \simeq n_\text{atom} X_0  \int d E_\gamma \frac{d N_\gamma}{d E_\gamma}  \int\limits_{\text{geom.}} d \cos{\theta_a}  \frac{d \sigma}{d \cos{\theta}_a}  \left( e^{- z_\text{min}  (m_a / p_z^a)  \Gamma_a} - e^{- z_\text{max}  (m_a / p_z^a)  \Gamma_a} \right) \Theta(p_a - \gamma_a^{(\text{min})} \, m_a)
~.
\end{align}
\end{widetext}
In the above expression, most variables are defined as explained below Eq.~(\ref{eq:NsignalMuon}). $X_0 \simeq 1.76 \cm$ is the radiation length of iron, $dN_\gamma / d E_\gamma$ is the total number of secondary photons (simulated in {\tt PYTHIA}) per energy bin within the first radiation length, and $d \sigma / d \cos{\theta_a}$ is the differential cross-section for $\gamma A \to a A$ in the fixed-target (lab) frame, where $\theta_a$ measures the angle of the emitted ALP with respect to the beam-axis. In integrating over $\cos{\theta_a}$, we only include regions in which the $a$ is forward and within the geometric acceptance of the SeaQuest spectrometer. Furthermore, the minimum ALP boost, $\gamma_a^\text{(min)}$, in the Heaviside step-function of Eq.~(\ref{eq:NsignalALP}) guarantees that the photons from the decay $a \to \gamma \gamma$ also remain within the instrument. Note that in the fixed-target frame, the ALP momentum, $p_a$, is a non-trivial function of $\cos{\theta_a}$. 

In Fig.~\ref{fig:ALP}, we show the projected SeaQuest reach both at Phase I (solid) and Phase II (dashed) for the  $7 \m - 8 \m$ fiducial decay region. We also compare the sensitivity of SeaQuest to existing constraints (gray) (see, e.g., Refs.~\cite{Jaeckel:2015jla,Dobrich:2015jyk,Dolan:2017osp} for a review) and other upcoming and proposed searches such as Belle-II (green)~\cite{Dolan:2017osp}, a beam dump run at NA62 (cyan)~\cite{na62cern}, and the proposed SHiP experiment (red)~\cite{Dobrich:2015jyk}. As in the previous sections, SeaQuest is sensitive to larger couplings compared to longer baseline experiments. Assuming negligible background, Phase II of a SeaQuest-like setup is capable of exploring currently viable parameter space for ALP masses of $10 \MeV \lesssim m_a \lesssim \text{GeV}$.

\section{Conclusion}\label{sec:conclusions}

The SeaQuest spectrometer is currently operating at Fermilab with access to the 120 GeV main injector proton beam. It is a nuclear physics experiment designed to measure Drell-Yan production of muons that originate from the collisions of the proton beam with various nuclear and polarized targets. Most of the proton beam remains unattenuated by the thin nuclear target and is dumped downstream onto a thick iron magnet, from which a large flux of exotic light and feebly interacting particles might be produced. If such particles are sufficiently long-lived and decay back to Standard Model species, the existing SeaQuest spectrometer can be leveraged to search for energetic leptons, hadrons, or photons reconstructing displaced vertices. 

A displaced vertex trigger has recently been installed in order to search for muons originating from the decays of long-lived and low-mass particles and is expected to acquire $\sim 10^{18}$ protons on target parasitically during the next two years. In this study, we have focused on a planned upgrade to install a recycled electromagnetic calorimeter, which would allow for a nearly background-free search for displaced electrons. There are improvements planned for the Fermilab accelerator complex in the coming years, which aim at providing a proton beam power capability of at least 1 megawatt for the DUNE program~\cite{Lebedev:2017vnu}. The Phase II luminosity and resulting gains in new physics sensitivity discussed in this work could be achieved by diverting a few percent of this beam directly to SeaQuest on a year timescale.

In this study, we have discussed signals arising from various benchmark scenarios such as minimal models of dark photons and dark Higgs bosons, inelastic dark matter, leptophilic scalars, and axion-like particles. A similar investigation within the context of strongly interacting dark matter was also presented in Ref.~\cite{Berlin:2018tvf}. The projected sensitivity of SeaQuest is comparable and complementary to future runs at NA62 as well as other proposed futuristic experiments such as FASER and SHiP. The fact that this discovery potential is possible with minimal modifications to the existing instrument provides an exciting opportunity. This warrants more dedicated detector simulations in order to fully optimize the instrumental layout. In future work, it would also be interesting to pursue more detailed calculations of signal yield for leptophilic scalars, dark Higgs bosons, axion-like particles, and other models not considered in this paper, such as GeV-scale sterile neutrinos. The SeaQuest experiment has the potential to discover a wide variety of dark sector models, which motivates a broad physics program at Fermilab in the search for new physics.

\hspace{-0.7cm}


\section*{Acknowledgements}
We would like to thank Nikita Blinov, Patrick deNiverville, Bertrand Echenard, Jared Evans, Ramona Grober, Eder Izaguirre, Yoni Kahn, Felix Kling, Gordan Krnjaic, Ming Liu, Kun Liu, Maxim Pospelov, Brian Shuve, Mike Williams, Sho Uemura, and Yiming Zhong for useful discussions. AB, PS, and NT are supported by the U.S. Department of Energy under Contract No. DE-AC02-76SF00515. SG acknowledges support from the University of Cincinnati. SG is supported by the NSF CAREER grant PHY-1654502. SG is grateful to the hospitality of both the Kavli Institute for Theoretical Physics in Santa Barbara, CA, supported in part by the National Science Foundation under Grant No. NSF PHY11-25915, the Aspen Center for Physics, supported by the National Science Foundation Grant No. PHY-1066293, and the Mainz Institute for Theoretical Physics (MITP) where some of the research reported in this work was carried out. 
\bibliography{SeaQuest}

\end{document}